\documentclass[a4paper,11pt]{article}
\pdfoutput=1 

\usepackage{jcappub} 

\usepackage[T1]{fontenc} 

\usepackage{url}
\usepackage{color}
\usepackage{comment}
\usepackage{mathrsfs}
\usepackage{hyperref}
\usepackage{braket}
\usepackage[dvipsnames]{xcolor}
\usepackage{natbib}
\usepackage{physics}
\usepackage{calc}
\usepackage{amsmath, amssymb}

\usepackage{here}
\usepackage{siunitx}
\usepackage{float}
\usepackage{url}
\usepackage{latexsym}
\usepackage{listings, jvlisting}
\usepackage{ascmac}
\usepackage{mathcomp}

\usepackage{minitoc}
\usepackage{amsmath}
\usepackage{amssymb} 

\usepackage{comment}
\usepackage{bm}

\newcommand{\simgt}{\lower.5ex\hbox{$\; \buildrel > \over \sim \;$}}

\title{{Analytical method for computing the covariance matrix 
of cosmic shear two-point correlation function}}

\author[a,b]{Kosuke~Nagura,}
\author[c,a]{Ryo~Terasawa,}
\author[a,b]{Taisei~Terawaki,}
\author[a,d]{Masahiro~Takada,}

\affiliation[a]{Kavli Institute for the Physics and Mathematics of the Universe (WPI), The University of Tokyo Institutes for Advanced Study (UTIAS), The University of Tokyo, 5-1-5 Kashi- wanoha, Kashiwa-shi, Chiba, 277-8583, Japan}
\affiliation[b]{Department of Physics, The University of Tokyo, Bunkyo, Tokyo 113-0031, Japan}
\affiliation[c]{Center for Frontier Science, Chiba University, 1-33 Yayoi-cho, Inage-ku, Chiba 263-8522, Japan}
\affiliation[d]{Center for Data-Driven Discovery (CD3), Kavli IPMU (WPI), UTIAS, The University of Tokyo, Kashiwa, Chiba 277-8583, Japan}

\emailAdd{kosuke.nagura@ipmu.jp}

\emailAdd{ryo.terasawa@chiba-u.jp}

\emailAdd{taisei.terawaki@ipmu.jp}

\emailAdd{masahiro.takada@ipmu.jp}

\abstract{\\
Accurate estimation of the covariance matrix of cosmic shear statistics is essential for cosmological analyses using current and upcoming wide-area weak lensing surveys. In this work, we investigate analytical methods for computing the Gaussian covariance matrix of the cosmic shear two-point correlation function (2PCF), taking into account the effects of finite survey geometry. 
{We compute the covariance of 2PCF based on the improved Narrow Kernel Approximation (iNKA), with a projection using the Legendre transformation.}
We {also} consider other analytical covariance estimators,
the $f_{\mathrm{sky}}$ approximation and the weighted quartic-counts method.
We evaluate the accuracy of those analytical methods using the convergence fields with the HSC Year 3 survey mask as a test case.
We find that the covariance of the 2PCF obtained by 
{using} the iNKA does not reproduce the covariance measured directly from Gaussian simulations. Although the iNKA accurately models the diagonal structure of the harmonic-space covariance, residual inaccuracies in the off-diagonal components propagate through the Legendre transformation and significantly affect the real-space covariance. In contrast, the weighted quartic-counts method shows better agreement with the simulations.
Our results demonstrate that accurate modeling of the off-diagonal structure of the harmonic-space covariance is crucial for obtaining reliable covariance estimates of real-space weak lensing statistics in the presence of survey window effects.
}

\begin{document}
\maketitle

\section{Introduction} \label{sec:intro}

Current photometric surveys have given a wealth of information on the structure formation. 
When performing the cosmology analysis of such data, it is important to estimate the covariance accurately. In this work, we focus on the covariance of angular two-point correlation functions (2PCFs) and angular power spectrum, which are widely used in the literature.
Roughly speaking,
there are two ways to compute the covariance: to measure sample covariance from the many simulation realizations, or to model the covariance analytically.

The former one is used in the Hyper Suprime-Cam (HSC) analyses (e.g., \cite{HSC1_mock_Shirasaki2019,HSC3_cosmicShearReal, HSC3_cosmicShearFourier, 2025PhRvD.111f3509T, 2025arXiv251217022F}). By cutting out the survey region from full-sky simulations~\cite{raytracingTakahashi2017}, we can take into account for the effect of the survey geometry including the Super-Sample Covariance (SSC) (e.g.,~\cite{2013PhRvD..87l3504T,2022PhRvD.106h3504T, 2024PhRvD.109f3504T}). 
Being made off of N-body simulations, covariance from the full-sky simulations incorporate non-Gaussian Covariance which arise from connected trispectrum. For the surveys with relatively small sky area such as HSC, we can cut out multiple pseudo-independent realizations from one full-sky simulation, but for the {Stage-IV} surveys with area $\gtrsim 10,000~{\rm deg}^2$, only few realizations can be extracted from a full-sky simulation and making many full-sky simulations will be computationally expensive.

To circumvent the expensive simulations, modeling covariance analytically (e.g., \cite{2011ApJ...734...76S, 2017MNRAS.470.2100K,2025A&A...699A.124R,2019MNRAS.486...52S,Friedrich2020,2016MNRAS.456.2662F,2008A&A...477...43J,2018MNRAS.481.1337H, 2020MNRAS.497.2699F}) would be realistic option for the {Stage-IV} surveys.
In the analytic model, the covariance is calculated as a sum of Gaussian Covariance term and non-Gaussian Covariance term.
In this work, we focus on the Gaussian term which is dominant on the large scales.
In the upcoming survey with wide area, the large-scale data becomes  important~\cite{2025OJAp....8E..19T, 2026arXiv260310113D} and so is its covariance.

Survey masks introduce mode-coupling and affect covariance. The Gaussian Covariance for pseudo-$C_{\ell}$ estimator in the presence of mask is studied in Refs~\cite{2004MNRAS.349..603E, 2019JCAP...11..043G, pseudoCl_Nicola2021}.
For the two-point correlation functions, the effect of window can be taken into account by accurately calculating number of pairs of galaxies in angular bin for the shot/shape noise term~\citep{2018MNRAS.479.4998T}. For the mixed contributions from the sample variance and shot/shape noise (``mixed term"), triplet counting gives accurate estimate~\citep{2025A&A...699A.124R}.
For the sample variance-only term, corresponding calculation involves with quartic counts~\citep{2002A&A...396....1S, 2019MNRAS.486...52S} which can be computationally expensive. 
Hence, in this paper, focusing on the sample variance-only term, we explore the other approach where the real-space covariance is calculated as a double Legendre transform of harmonic space covariance with survey window effect. 
We will compare the both approach with the sample covariance measured from Gaussian realizations.

We organize this paper as follows. In Section~\ref{sec:Formulation}, we summarize how to calculate the covariance of 2PCFs and power spectra analytically.
In Section~\ref{sec:simulations}, we introduce Gaussian simulations of cosmic shear fields to measure the sample covariance with which we compare our analytical covariance estimation.
In Section~\ref{sec:results}, we present the accuracy of analytic covariance estimate compared to the sample covariance. Section~\ref{sec:conclusion} is devoted to discussion and conclusion.

\section{{Formulation of the covariance matrix of cosmic shear 
two-point correlation function}}\label{sec:Formulation}
\label{sec:covariance}

\subsection{{Power spectrum with and without survey window effects}}
\label{ssec:windowed_cell}

In this work, we consider the Gaussian covariance of weak gravitational lensing observables in a finite survey area. Specifically, we focus on the power spectrum, $C_{\ell}$,
of {the spin-0 
convergence field $\kappa(\hat{\bm{n}})$, defined on the celestial sphere.}
{In practice, a survey is limited to 
a partial sky coverage and has complex geometry involved with masks. 
These affect the measured power spectrum and its covariance.
We refer to this effect as the survey window effect and  
denote 
physical quantities affected by this effect with a tilde symbol, e.g.,
$\tilde{C}_{\ell}$.
}
We also consider the real-space counterpart of $C_\ell$, the two-point correlation function $\xi(\theta)$. 
{For clarity of presentation, we restrict ourselves to the spin-$0$ field $\kappa$ in the main text, and present the corresponding equations and results for the spin-2 shear fields, which are more direct weak lensing observables in the Appendix.}
For simplicity, we restrict our analysis to a single {source redshift}.

On the celestial sphere, the convergence field 
$\kappa$ {in the direction $\hat{\bm{n}}$} is expanded in spherical harmonics $Y_{\ell m}$ as
\begin{align}
    \kappa(\hat{\bm{n}}) = \sum_{\ell,m} a_{\ell m} Y_{\ell m}(\hat{\bm{n}}),
\end{align}
and the power spectrum is defined by
\begin{align}
    \expval{a_{\ell m}a_{\ell' m'}^*}
    =
    \delta^K_{\ell\ell'}
    \delta^K_{mm'}
    C_{\ell},
\end{align}
{where $\delta^K_{\ell\ell'}$ is the Kronecker delta function.}

The observed field including 
{the survey window effect}
{can be expressed}
as
\begin{align}
    \tilde{\kappa}(\hat{\bm{n}})
    =w(\hat{\bm{n}})\kappa(\hat{\bm{n}}),
\end{align}
{where $w(\hat{\bm{n}})$ is the survey mask field, defined as $w(\hat{\bm{n}})=1$ if $\hat{\bm{n}}$ lies inside
the survey region, and $w(\hat{\bm{n}})=0$ otherwise. When weights are also taken into account, the following discussion can be straightforwardly extended.}
The resulting power spectrum can be expressed using the mode-coupling matrix 
${\cal M}_{\ell\ell'}(w,w)$ 
\citep{2011MNRAS.412...65H,2019MNRAS.484.4127A,pseudoCl_Nicola2021}
as
\begin{align} \label{eq:cell_to_tildecell}
    \tilde{C}_{\ell} = \sum_{\ell'} {\cal M}_{\ell\ell'}(w,w) C_{\ell'}.
\end{align}
{From this equation, we can obtain the unwindowed (i.e.,  window-corrected)
power spectrum via $C_{\ell}={\cal M}^{-1}_{\ell\ell'}\tilde{C}_{\ell'}$.
This is the so-called pseudo-$C_\ell$ method for estimating $C_\ell$.}

\subsection{{Relation between $\xi (\theta)$ and $\tilde{C}_\ell$}}
\label{ssec:xi_windowed_cell}
The two-point correlation function of the convergence field is given by the Legendre transform of the power spectrum as
\begin{align}
    \xi(\theta)
    =\sum_{\ell}\frac{2\ell+1}{4\pi}C_{\ell}P_{\ell}(\cos\theta),
\end{align}
where $P_\ell(x)$ is the Legendre polynomial.

{The measured correlation function is an unbiased estimator of the underlying correlation function, in the sense that it is
not affected by the survey window.}
Nevertheless, Ref.~\cite{Friedrich2020} derived the relation between the windowed (measured) power spectrum and the correlation function (see Appendix~C {of Ref.~\cite{Friedrich2020}}):
\begin{align}
    \xi_i
    = \frac{n^2}{N_{\mathrm{pair},i}} \tilde{\xi}_i,
    \label{eq:xi_window-xi_relation_binned}
\end{align}
where $\xi_i \equiv \xi[\theta_{i,-}, \theta_{i,+}]$ denotes the correlation function averaged over the $i$-th angular bin, $n$ is the mean number density of source galaxies per steradian, and $N_{\mathrm{pair},i}$ is the number of galaxy pairs in the $i$-th bin within the finite survey region. 
The quantity $\tilde{\xi}_i$ is the ``window-affected'' correlation function, defined in terms of the windowed power spectrum $\tilde{C}_\ell$ as
\begin{align}
    \tilde{\xi}_i
    = 2\pi \sum_{\ell}
    \left[
    P_{\ell+1}(x) - P_{\ell-1}(x)
    \right]_{\cos\theta_{i,+}}^{\cos\theta_{i,-}}
    \tilde{C}_{\ell},
    \label{eq:tildexi_tildecell_binned}
\end{align}
{where we took average of $P_{\ell}(\cos\theta)$ over the angular bin using recursion relation of Legendre polynomials.}
Eq.~(\ref{eq:xi_window-xi_relation_binned}) is non-trivial, and we will validate it using simulated maps of the weak lensing field.

\subsection{Covariance matrix of $\xi(\theta)$}

{Following Appendix~C of Ref.~\cite{Friedrich2020} and as indicated by Eqs.~(\ref{eq:xi_window-xi_relation_binned}) and (\ref{eq:tildexi_tildecell_binned}),} the covariance matrix of the two-point correlation function can be {\it formally}
expressed in terms of the covariance of $\tilde{C}_\ell$ as
\begin{align} 
    \mathrm{Cov}
    (\xi_i,\xi_j)
    &=
    \frac
    {n^4}
    {N_{\mathrm{pair},i}N_{\mathrm{pair},j}}
    \notag\\
    &\times(2\pi)^2
    \sum_{\ell,\ell'}
    [P_{\ell+1}(x)-P_{\ell-1}(x)]_{\cos\theta_{i,+}}^{\cos\theta_{i,-}}
    [P_{\ell'+1}(x)-P_{\ell'-1}(x)]_{\cos\theta_{j,+}}^{\cos\theta_{j,-}}
    \times\mathrm{Cov}(\tilde{C}_\ell,\tilde{C}_{\ell'}),
    \label{eq:maskedCellCovariance_projection}
\end{align}
{where ${\rm Cov}(\tilde{C}_\ell,\tilde{C}_{\ell'})$ is the covariance matrix of $\tilde{C}_\ell$. As explicitly given by Eq.~(C12) of \cite{Friedrich2020}, this is given in terms of the underlying power spectrum $C_\ell$ and the window function. Computing ${\rm Cov}(\xi_i,\xi_j)$ directly requires 
$O(\ell_{\rm max}^6)$ operations and is therefore computationally intractable,
since we are interested in angular scales up to $\ell_{\rm max}\sim \mbox{a few}\times O(10^3)$ (or even $O(10^4))$. For this reason, we need to adopt an approximation for ${\rm Cov}(\tilde{C}_\ell,\tilde{C}_{\ell'})$ if we want to 
proceed with an analytical computation of ${\rm Cov}(\xi_i,\xi_j)$, as proposed 
in Refs.~\citep{2004MNRAS.349..603E,2019JCAP...11..043G,pseudoCl_Nicola2021}.
}

\subsubsection{Improved Narrow Kernel Approximation (iNKA)}
{{In this paper,}
we adopt the improved Narrow Kernel Approximation (iNKA) \cite{pseudoCl_Nicola2021} 
to calculate the covariance matrix ${\rm Cov}(\tilde{C}_\ell,\tilde{C}_{\ell'})$ 
(see also \cite{2019JCAP...11..043G} for the original NKA formalism).}
{The iNKA} assumes that the mode-coupling matrix is sharply localized around $\ell \simeq \ell'$, thereby enabling a fast computation of the Gaussian covariance of $\tilde{C}_\ell$:
\begin{align}\label{eq:iNKA}
    \mathrm{Cov}(\tilde{C}_\ell,\tilde{C}_{\ell'})
    \simeq
    \frac{1}{2}
    \left(
    \frac
    {\tilde{C}_\ell+\tilde{C}_{\ell'}}
    {\expval{w^2}_{\mathrm{pixel}}}
    \right)^2
    \Xi_{\ell\ell'}(w^2,w^2),
\end{align}
where $\langle~ \cdot~ \rangle_{\rm pix}$ denotes the averaging the quantity 
inside brackets over all pixels, and 
the matrix
$\Xi_{\ell\ell'}(w^2,w^2)$ is defined as
\begin{align}
    \Xi_{\ell\ell'}(w^2,w^2)
\equiv 
    \frac{{\cal M}_{\ell\ell'}(w^2,w^2)}{2\ell'+1}.
\end{align}
{Here ${\cal M}_{\ell\ell'}$ is the mode-coupling matrix defined in a similar way to ${\cal M}_{\ell\ell'}$
Eq.~(\ref{eq:cell_to_tildecell}), but using the squared weight field $w^2$
instead of $w$ in the calculation.
The matrix $\Xi_{\ell\ell'}$ depends only on the weight field, and its computation is inexpensive when using a spherical harmonic expansion of the weight field.}

By substituting this expression into the covariance formula of 
${\rm Cov}(\xi_i,\xi_j)$ (Eq.~\ref{eq:maskedCellCovariance_projection}),
one can perform the calculation efficiently. In this work, we 
{assess}
the accuracy of this {analytical} approach 
by comparing {its predictions with those computed from realizations of simulated lensing 
fields.}

\subsubsection{The $f_\mathrm{sky}$ approximation}

{The $f_{\rm sky}$
approximation is commonly used in previous studies to 
compute the covariance of cosmic shear fields \cite{2020MNRAS.497.2699F}.}
The covariance of the power spectrum {with the $f_{\rm sky}$ approximation} is given by
\begin{align}
    \mathrm{Cov}
    (C_\ell, C_{\ell'})^{f_\mathrm{sky}}
    &=
    \delta^K_{\ell \ell'}\frac{2C_\ell^2}{(2\ell+1)f_\mathrm{sky}},
\end{align}
where 
$f_{\rm sky}$ is the sky fraction covered by the survey.
The covariance of $\xi$ can be computed by 
{integrating} the power spectrum covariance, weighted by Legendre polynomials, over the bin width:
\begin{align}
    \mathrm{Cov}
    (\xi_i,\xi_j)^{f_\mathrm{sky}}
    &=
    \sum_{\ell, \ell'}
    \frac
    {[P_{\ell+1}(x)-P_{\ell-1}(x)]_{\cos\theta_{i,+}}^{\cos\theta_{i,-}}
    [P_{\ell'+1}(x)-P_{\ell'-1}(x)]_{\cos\theta_{j,+}}^{\cos\theta_{j,-}}}
    {(4\pi)^2(\cos\theta_{i,-}-\cos\theta_{i,+})(\cos\theta_{j,-}-\cos\theta_{j,+})}
    \notag \\
    &\times\mathrm{Cov}
    (C_\ell, C_{\ell'})^{f_\mathrm{sky}}.
\end{align}
We will compare the covariance obtained with the $f_{\rm sky}$ approximation 
to those obtained from the simulations and the iNKA. 

\section{Gaussian simulations of cosmic shear fields}
\label{sec:simulations}

\subsection{{Cosmological parameters and simulation setups}}

{To assess the accuracy of the analytical methods for computing the Gaussian
covariance matrix of cosmic shear correlation functions, we use}
a large number of Gaussian realizations {of cosmic shear fields.
In this subsection, we describe the method used to construct the Gaussian simulations.}

First, we 
{compute}
the input power spectrum {of the convergence field for a fiducial flat $\Lambda$CDM model}
using the Core Cosmology Library (\texttt{CCL})~\cite{2019ApJS..242....2C}. 
{The flat $\Lambda$CDM model is specified by}
the following parameters:
$\Omega_c = 0.233$ and  
$\Omega_b = 0.046$, the density parameters of CDM and baryon, respectively; 
$\sigma_8 = 0.82$ and  
$n_s = 0.97$, the normalization and spectral index of the linear power spectrum; and the Hubble parameter
$h = 0.7$.
{We use \texttt{HEALPix} \cite{2005ApJ...622..759G} to generate the convergence 
fields from the input power spectrum on the celestial sphere.
Here we adopt a pixel resolution of $N_{\rm side}=2048$, corresponding to 
a pixel scale of approximately 1.7~arcmin.
Then we apply the HSC Year 3 (HSC-Y3) survey mask \cite{HSCY3_catalog_Li22} to obtain a masked map of the lensing fields. We use 1,000 realizations of the lensing fields for statistical analyses.}

\subsection{Measurements of $\tilde{C}_\ell$, $\xi(\theta)$, 
and their covariance matrices from simulated maps}

{From}
each {convergence map,}
we 
{measure}
$\tilde{C}_\ell$, which is affected by the survey window,  
using \texttt{NaMaster} \cite{2019MNRAS.484.4127A}.
We {compute}
the mean and covariance over the realizations as follows:
\begin{align}
    &\tilde{C}_\ell^{\mathrm{sim}} 
    = 
    \frac{1}{N_{\mathrm{sim}}}\sum_{r=1}^{N_\mathrm{sim}}
    \tilde{C}_\ell^{(r)},
    \nonumber \\
    &\mathrm{Cov}
    (\tilde{C}_\ell, \tilde{C}_{\ell'})^{\mathrm{sim}}
    =
    \frac{1}{N_\mathrm{sim}-1}
    \sum_{r=1}^{N_\mathrm{sim}}
    (\tilde{C}_\ell^{(r)}-\tilde{C}_\ell^{\mathrm{sim}})
    (\tilde{C}_{\ell'}^{(r)}-\tilde{C}_{\ell'}^{\mathrm{sim}}),
    \label{eq:cov_sim_def}
\end{align}
{where $N_{\rm sim}=1,000$, the number of the Gaussian realizations.}

For each realization, we construct a convergence catalog by assigning to each galaxy position in the HSC-Y3 galaxy catalog the value of the convergence field at the corresponding position on the masked map {pixel}. We then measured the two-point correlation function from the resulting catalog using \texttt{TreeCorr} \cite{treecorr}. Repeating this procedure for all realizations, we estimated the mean and covariance over the realizations in the same manner as for the power spectrum:
\begin{align}
    \xi^{\mathrm{sim}}_i
    &=
    \frac{1}{N_{\mathrm{sim}}}\sum_{r=1}^{N_\mathrm{sim}}
    \xi^{(r)}_i
    \nonumber \\
    \mathrm{Cov}
    (\xi_i,\xi_j)^{\mathrm{sim}}
    &=
    \frac{1}{N_\mathrm{sim}-1}
    \sum_{r=1}^{N_\mathrm{sim}}
    (\xi^{(r)}_i- \xi^{\mathrm{sim}}_i)
    (\xi^{(r)}_j- \xi^{\mathrm{sim}}_j).
\end{align}
We employ 20~logarithmically-spaced bins in the range $\theta=[6,300]~$arcmin.

\section{Results}\label{sec:results}

\subsection{{Signal of $\tilde{C}_{\ell},~\xi(\theta)$}}

\begin{figure}
\centering

\begin{minipage}[t]{.48\textwidth}
\vspace{0pt}
\centering
\includegraphics[width=\linewidth]{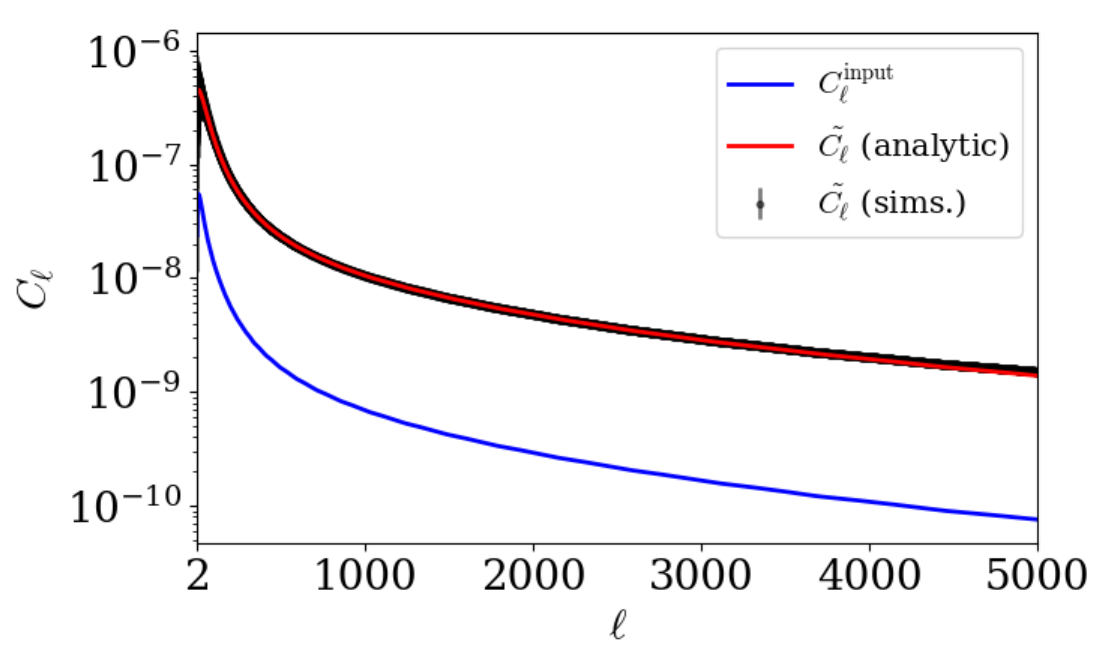}
\caption{Input power spectrum $C_\ell$ (blue line), the window-convolved power spectrum $\tilde{C}_\ell$ obtained via the mode-coupling matrix (red line), and the mean power spectrum measured from Gaussian simulations (black points with error bars). The agreement demonstrates that the mode-coupling matrix accurately captures the survey window effect.}
\label{fig:tildeCl_to_Cl}
\end{minipage}
\hfill
\begin{minipage}[t]{.48\textwidth}
\vspace{0pt}
\centering
\includegraphics[width=\linewidth]{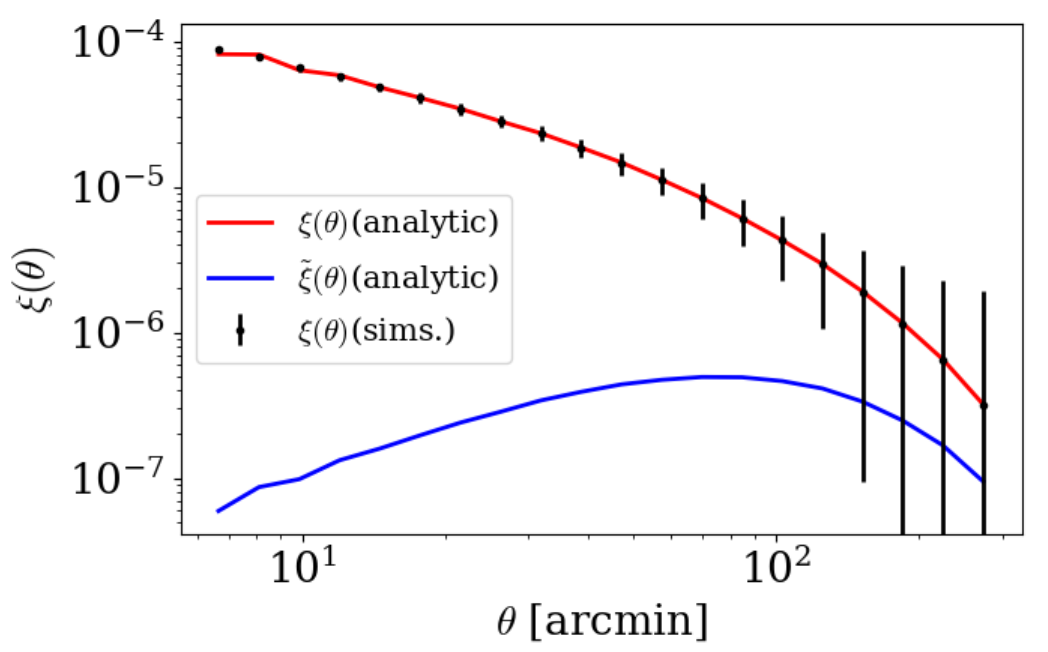}
\caption{Mean of correlation function $\xi^{\rm sim}$ (black points with error bars). Error bars correspond to $\sqrt{{\rm Cov} (\xi, \xi)^{\rm sim}}$. 
Blue line shows $\tilde{\xi}$ calculated via Eq.~\ref{eq:tildexi_tildecell_binned}. 
Red line shows $\xi$ calculated via Eq.~\ref{eq:xi_window-xi_relation_binned}.
}
\label{fig:tildexi_to_xi}
\end{minipage}

\end{figure}

First, we validate the relation between $\xi$ and $\tilde{\xi}$ (Eq.~\ref{eq:xi_window-xi_relation_binned}) which provides the basis for the following discussion on the covariance of $\xi$.

Fig.~\ref{fig:tildeCl_to_Cl} shows the mean of the measured power spectrum, $\tilde{C}_{\ell}^{\rm sim}$ (black points), the input theoretical power spectrum $C_\ell$ (blue line), and the window-convolved power spectrum $\tilde{C}_\ell$ obtained via the mode-coupling matrix (red line). We can see that the multiplication of the mode-coupling matrix (Eq.~\ref{eq:cell_to_tildecell}) captures the survey window effect accurately.

We can correct the window effect 
on correlation function by multiplying the corresponding $\theta$-dependent factor as in Eq.~\ref{eq:xi_window-xi_relation_binned}.
Fig.~\ref{fig:tildexi_to_xi} shows the measured mean of correlation function $\xi^{\rm sim}$ (black points with errorbars).
First, we calculate $\tilde{\xi}$ (blue line) using $\tilde{C_{\ell}}$ via Eq.~\ref{eq:tildexi_tildecell_binned}. Then we obtain ${\xi}$ (red line) using Eq.~\ref{eq:xi_window-xi_relation_binned}.
We find good agreement between $\xi$ (Eq.~\ref{eq:xi_window-xi_relation_binned}) and measured one. 

\subsection{Covariance of $\xi(\theta)$}

\begin{figure}
\centering
\includegraphics[width=.9\linewidth]{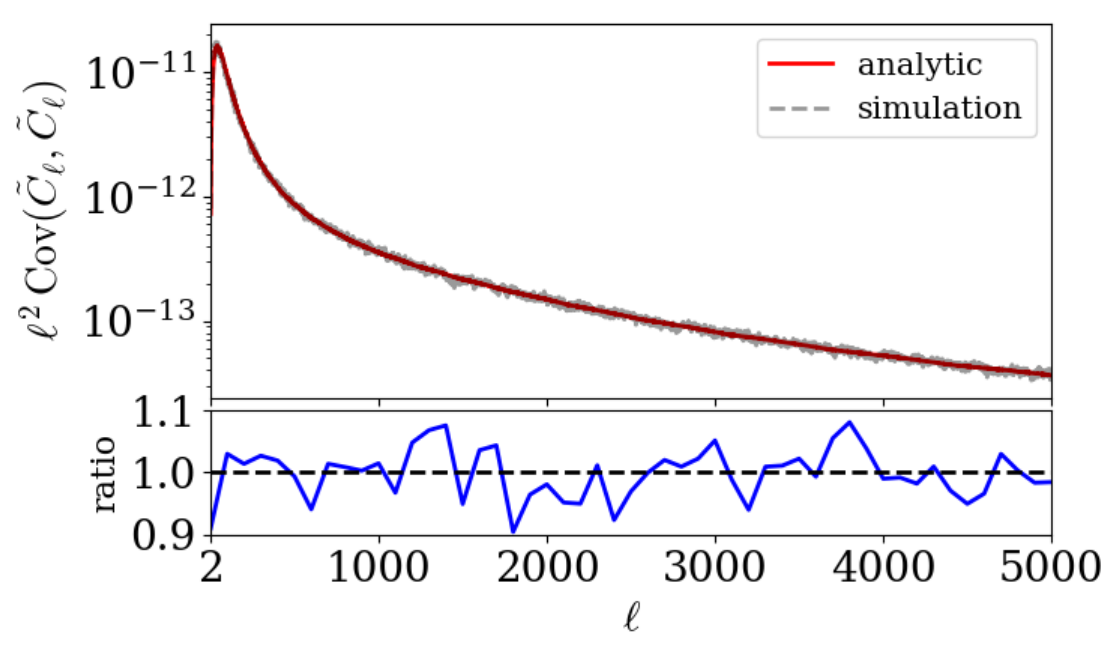}
\caption{
Upper panel: Diagonal components of
$\mathrm{Cov}(\tilde{C}_\ell,\tilde{C}_{\ell'})$,
measured from {1000} Gaussian realizations (dashed line)
and calculated using Eq.~\ref{eq:iNKA} (solid lines).
Lower panel: Ratio of the analytical prediction
based on the iNKA to the simulation results.
The ratio is consistent within~10\%.
}
\label{fig:cov_cell}
\end{figure}

First, we compare the analytically calculated covariance
$ \mathrm{Cov}(\tilde{C}_\ell,\tilde{C}_{\ell'})$
based on iNKA with the covariance measured from the simulations,
$\mathrm{Cov}(\tilde{C}_\ell, \tilde{C}_{\ell'})^{\mathrm{sim}}$.
Fig.~\ref{fig:cov_cell} shows that, for the diagonal components, we find good agreement between the two. The ratio between simulations and iNKA is consistent within~10\%.
\begin{figure}
    \centering
    \includegraphics[width=130mm]{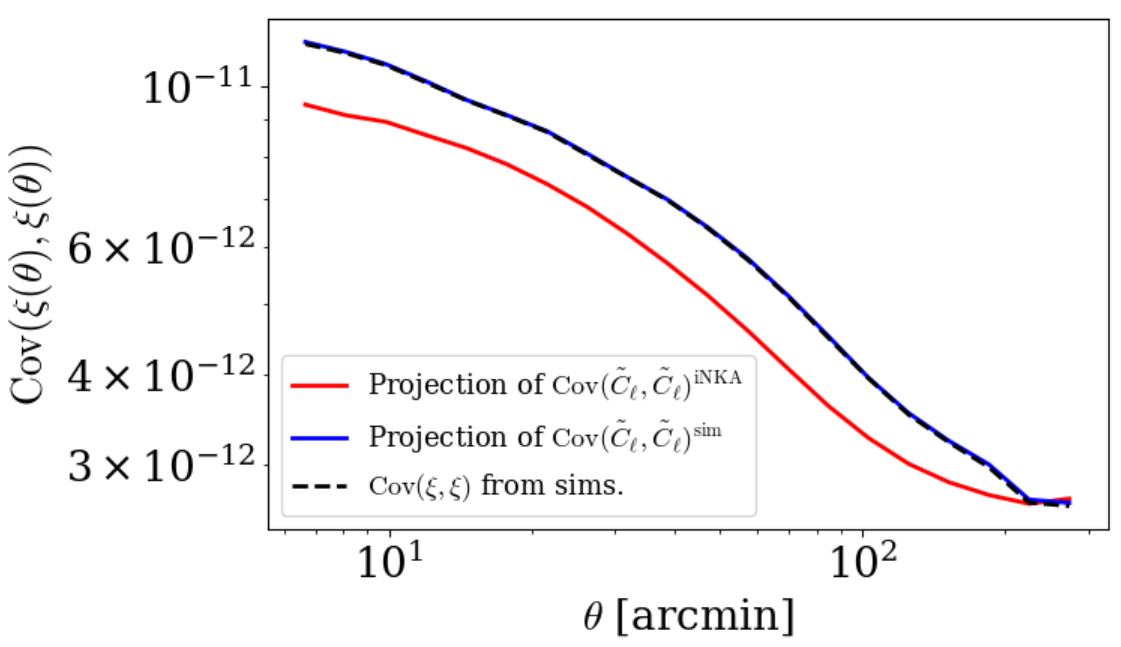}
    \caption{Diagonal components of $\mathrm{Cov}(\xi[\theta_-,\theta_+],\xi[\theta'_-, \theta'_+])$, measured from Gaussian realizations (dashed line) and calculated using Eq.~\ref{eq:maskedCellCovariance_projection} (solid lines).
    }
    \label{fig:cov_xi_kappa_projection}
\end{figure}

Fig.~\ref{fig:cov_xi_kappa_projection} shows diagonal components of $\mathrm{Cov}(\xi_i,\xi_j)$, measured from Gaussian realizations (dashed line) and calculated using Eq.~\ref{eq:maskedCellCovariance_projection} (solid lines). In Eq.~\ref{eq:maskedCellCovariance_projection}, we set $\ell_{\rm max} = 3N_{\rm{side}}-1 = 6143$. The good agreement between projection of $\mathrm{Cov}(\tilde{C}_\ell,\tilde{C}_{\ell'})$ measured from Gaussian simulations (blue) and $\mathrm{Cov}(\xi_i,\xi_j)^{\rm sim}$ indicates Eq.~\ref{eq:maskedCellCovariance_projection} is accurate and taking $\ell_{\rm max} = 6143$ is enough to calculate the covariance accurately. On the other hand, projection of $\mathrm{Cov}(\tilde{C}_\ell,\tilde{C}_{\ell'})$ calculated using iNKA (red) fails to reproduce $\mathrm{Cov}(\xi_i,\xi_j)^{\rm sim}$. 

A possible reason why
$\mathrm{Cov}(\xi_i,\xi_j)$
computed using $\mathrm{Cov}(\tilde{C}_\ell,\tilde{C}_{\ell'})^{\rm iNKA}$ does not agree with
$\mathrm{Cov}(\xi_i,\xi_j)^{\mathrm{sim}}$,
measured directly from Gaussian simulations, is the limited accuracy of the iNKA itself.

\begin{figure}
\centering
\includegraphics[width=.9\linewidth]{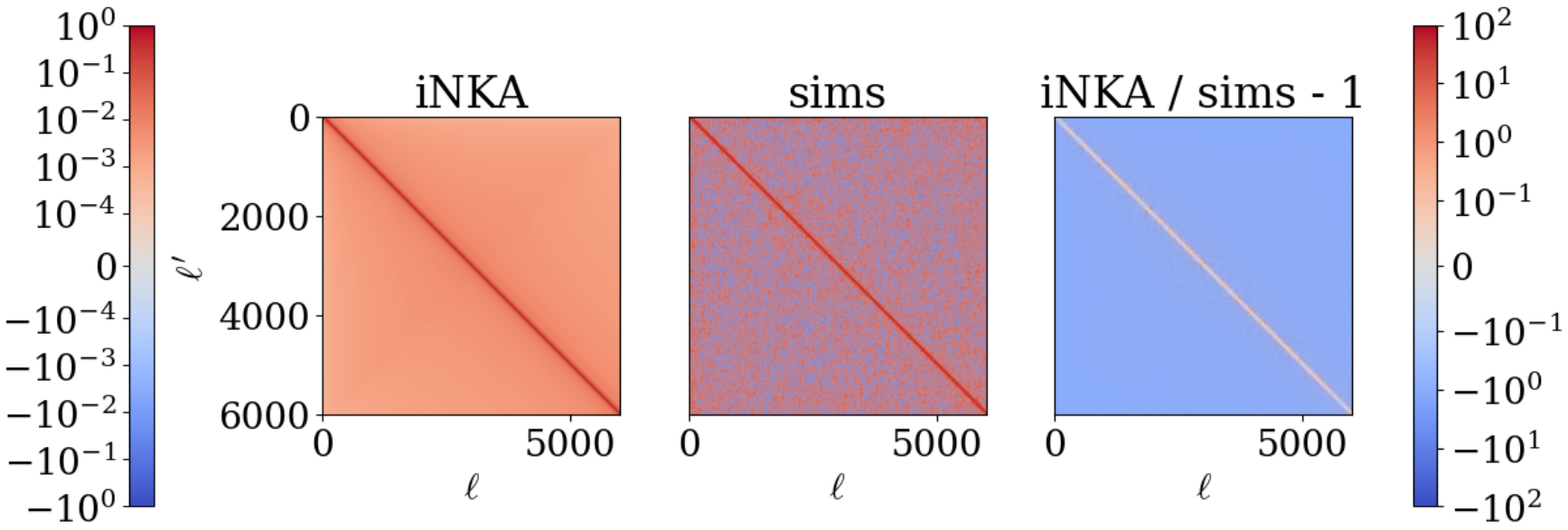}
\caption{Correlation matrices of $\mathrm{Cov}(\tilde{C}_\ell,\tilde{C}_{\ell'})$
computed using the iNKA (left) and measured from Gaussian simulations (center).
{The colors corresponding to the map values are depicted on the left.
The right plot shows the fractional difference $\mathrm{Cov}(\tilde{C}_\ell,\tilde{C}_{\ell'})^{\rm iNKA} / \mathrm{Cov}(\tilde{C}_\ell,\tilde{C}_{\ell'})^{\rm sim} - 1$ and color bar is shown on the right.}
The iNKA reproduces the diagonal structure accurately,
while noticeable differences remain in the off-diagonal components.}
\label{fig:matrix}
\end{figure}

Fig.~\ref{fig:matrix} compares the correlation matrices of
$\mathrm{Cov}(\tilde{C}_\ell,\tilde{C}_{\ell'})$
computed using the iNKA and measured from Gaussian simulations.
The right panel shows the ratio between the two covariance estimates.
Although the iNKA reproduces the diagonal structure accurately,
visible discrepancies remain in the off-diagonal components.
This behavior is consistent with the underlying assumption of iNKA. Specifically, the iNKA relies on the mode-coupling kernel being sharply localized around $\ell \simeq \ell'$. It is therefore effective for diagonal-dominated quantities, but its errors can be amplified in situations where off-diagonal information is required with high precision.
Indeed, in the calculation of
$\mathrm{Cov}(\xi_i,\xi_j)$,
the contribution from the off-diagonal components of
$\mathrm{Cov}(\tilde{C}_\ell,\tilde{C}_{\ell'})$
is not negligible
{as can be seen in Eq.~\ref{eq:maskedCellCovariance_projection}}.
Therefore, if one computes
$\mathrm{Cov}(\xi_i,\xi_j)$
by simply extending the iNKA, the resulting accuracy will be degraded as shown in Fig.~\ref{fig:cov_xi_kappa_projection}.

\subsection{{2PCF covariance from quartic counts}}
\label{ssec:shirasaki}

Alternatively, the 2PCFs covariance can be calculated as galaxy quartic counts weighted by $\xi \times \xi$ (and galaxy weights)~\citep{2002A&A...396....1S, 2019MNRAS.486...52S}:

\begin{align}
\label{eq:shirasaki_quartic}
    \text{Cov}({\xi}_{i},{\xi}_{j})&= \frac{1}{N_{{\rm pair}, i} N_{{\rm pair}, j}}
\sum_{a,b,c,d} w_a w_b w_c w_d 
\Delta_{\theta_i}(\theta_{ab})\Delta_{\theta_j}(\theta_{cd})
\Bigl\{
  \xi(\theta_{ac})\xi(\theta_{bd})
  + \xi(\theta_{ad})\xi(\theta_{bc}) 
\Bigr\}/2, \\ \nonumber
\end{align}
where $w_a$ is the measurement weight for the $a$-th galaxy, $\theta_{ab}$ is angular separation between the $a$-th and $b$-th galaxies, and 
$\Delta_{\theta_i}(\theta_{ab})$ is $1$ if $\theta_{i,-} < \theta_{ab} < \theta_{i,+}$ and $0$ otherwise.  
We have neglected the terms involved with the shape noise.
We derive the matrix expression of Eq~\eqref{eq:shirasaki_quartic} to make the calculation faster: 
\begin{align}
\text{Cov}({\xi}_{i},{\xi}_{j})
&=
\frac{1}{N_{{\rm pair}, i} N_{{\rm pair}, j}}\left[
\mathbf{w}_i^{\mathrm{T}}
\left(
\mathbf{\Xi}_{1,i}
\mathbf{\Xi}_{1,j}^{\mathrm{T}}
+
\mathbf{\Xi}_{2,i}
\mathbf{\Xi}_{2,j}^{\mathrm{T}}
\right)
\mathbf{w}_j
\right]/2 .
\label{eq:shirasaki_matrix_form}
\end{align}
The vector $\mathbf{w}_i$ ($\mathbf{w}_j$) with the length $N_{{\rm pair},i}$ ($N_{{\rm pair},j}$) is defined by arranging the pair weights $w_a w_b$ ($w_c w_d$).
The vector $\mathbf{\Xi}_{1,i}$ ($\mathbf{\Xi}_{1,j}$) with the length $N_{{\rm pair},i}$ ($N_{{\rm pair},j}$) is defined by arranging
$\xi({\theta_{ac})}$ ($\xi({\theta_{bd})}$),
and similarly for the other combinations.
{We further merge galaxies into coarse grid ($N_{\rm side} = 256$ or $512$ depending on $\theta_i$) to reduce the size of the matrices.}
In the following, we consider one of the six sub-fields in HSC-Y3, XMM field, 
and employ 10~logarithmically-spaced bins in the range $\theta=[6,300]~$arcmin,
to ease the computational expense. 

\begin{figure}
    \centering
    \includegraphics[width=130mm]{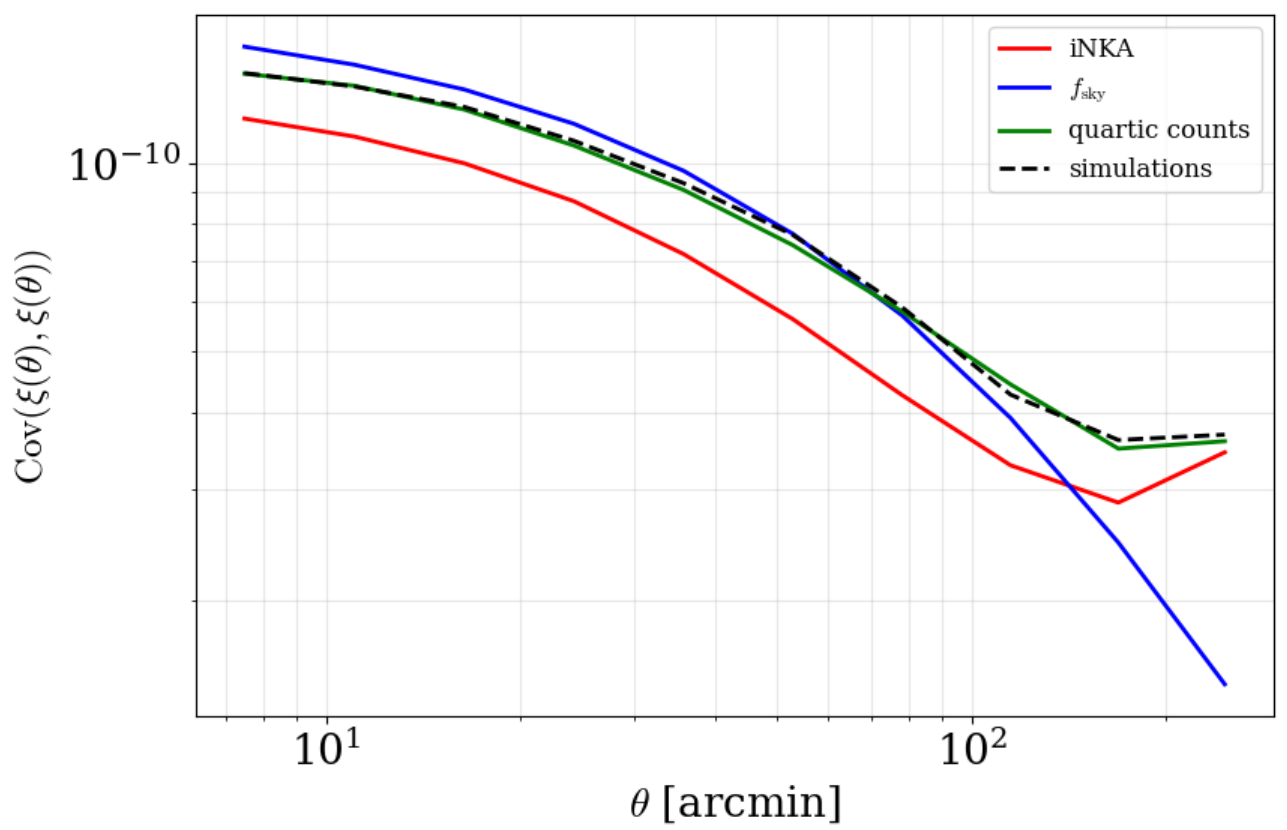}
    \caption{Diagonal components of $\mathrm{Cov}(\xi[\theta_-,\theta_+],\xi[\theta'_-, \theta'_+])$, measured from Gaussian realizations (dashed line) and calculated using analytical methods (solid lines).
    {Note that we consider XMM field only for this figure.}
    }
   \label{fig:cov_xi_kappa_analytical_methods}
\end{figure}

Figure~\ref{fig:cov_xi_kappa_analytical_methods} compares the diagonal components of the covariance matrix computed using three analytical methods with those measured from Gaussian simulations. 
We find that the quartic counts method reproduces the simulation results more accurately than the approaches based on the iNKA and $f_{\mathrm{sky}}$ approximations.

\section{Conclusion}\label{sec:conclusion}

In this work, we {have} investigated analytical methods for computing the covariance matrix of the cosmic shear two-point correlation function, taking into account the effects of the survey window.

Our approach is based on constructing the covariance of the real-space correlation function $\xi(\theta)$ from the covariance of the windowed power spectrum $\tilde{C}_\ell$, using the Legendre transformation. As a prerequisite, we {have} validated that the pseudo-$C_\ell$ framework correctly describes the impact of the survey window, and that the relation between $\tilde{\xi}$ and $\xi$ (Eq.~\ref{eq:xi_window-xi_relation_binned}) provides an accurate mapping between harmonic and real space.

The main result of this study is that the covariance of $\xi(\theta)$ obtained by projecting the covariance of $\tilde{C}_\ell$ computed under the iNKA fails to reproduce the covariance measured from Gaussian simulations, even at the level of the diagonal components. This discrepancy arises because the iNKA, while {it is} accurate for the diagonal components of $\mathrm{Cov}(\tilde{C}_\ell, \tilde{C}_{\ell'})$, does not adequately capture the off-diagonal structure. Since the computation of $\mathrm{Cov}(\xi(\theta), \xi(\theta'))$ involves contributions from off-diagonal modes through the Legendre transformation, the inaccuracy propagates and leads to a degradation in the final covariance.

We {have} compared this approach with other analytical methods, including the $f_{\mathrm{sky}}$ approximation and the weighted quartic counts method proposed by Schneider et al.~(2002)~\cite{2002A&A...396....1S}. We find that the quartic counts method provides better agreement with the simulation results, highlighting the importance of accurately modeling pair-based correlations in configuration space.

Our results demonstrate that achieving an accurate prediction of the covariance matrix in real space requires precise modeling of mode coupling beyond diagonal approximations in harmonic space. This has important implications for future wide-area surveys, where accurate covariance estimation is essential for cosmological parameter inference.

\appendix

\section{Covariance of spin-2 shear field}

\subsection{Relation between shear two-point correlation functions and the window-convolved power spectrum}
\label{app:xi_pseudo_cl_relation}

In this appendix, we derive the relation between the binned shear two-point
correlation functions and the window-convolved power spectrum.  

Let \(\hat{\bm n}_a\) and \(\hat{\bm n}_b\) be two directions on the sphere, and
let $ \hat{\bm n}_a \cdot \hat{\bm n}_b
    =
    \cos\theta_{ab}.$
We first use the completeness relation of the Wigner \(d\)-matrix,
\begin{equation}
    \delta_{\rm D}
    \left(
    \cos\theta_{ab}-\cos\theta
    \right)
    =
    \sum_{\ell}
    \frac{2\ell+1}{2}
    d^\ell_{ss'}(\theta)
    d^\ell_{ss'}(\theta_{ab}) .
    \label{eq:app_delta_wigner_d}
\end{equation}
The addition theorem for spin-weighted spherical harmonics gives
\begin{equation}
    \sum_{m=-\ell}^{\ell}
    {}_sY_{\ell m}^{*}(\hat{\bm n}_a)
    {}_{s'}Y_{\ell m}(\hat{\bm n}_b)
    =
    \frac{2\ell+1}{4\pi}
    D^\ell_{ss'}(\alpha,\theta_{ab},\gamma),
    \label{eq:app_spin_addition_theorem}
\end{equation}
where the Wigner \(D\)-matrix is written as
\begin{equation}
    D^\ell_{ss'}(\alpha,\theta_{ab},\gamma)
    =
    e^{-is\alpha}
    d^\ell_{ss'}(\theta_{ab})
    e^{-is'\gamma}.
\end{equation}
Equivalently,
\begin{equation}
    d^\ell_{ss'}(\theta_{ab})
    =
    \frac{4\pi}{2\ell+1}
    e^{is\alpha}
    e^{is'\gamma}
    \sum_{m=-\ell}^{\ell}
    {}_sY_{\ell m}^{*}(\hat{\bm n}_a)
    {}_{s'}Y_{\ell m}(\hat{\bm n}_b).
    \label{eq:app_d_to_spinY}
\end{equation}
Substituting Eq.~\eqref{eq:app_d_to_spinY} into
Eq.~\eqref{eq:app_delta_wigner_d}, we obtain
\begin{equation}
    \delta_{\rm D}
    \left(
    \cos\theta_{ab}-\cos\theta
    \right)
    =
    2\pi
    \sum_{\ell}
    d^\ell_{ss'}(\theta)
    e^{is\alpha}
    e^{is'\gamma}
    \sum_{m=-\ell}^{\ell}
    {}_sY_{\ell m}^{*}(\hat{\bm n}_a)
    {}_{s'}Y_{\ell m}(\hat{\bm n}_b).
    \label{eq:app_delta_spinY}
\end{equation}
We define the complex shear field as
\begin{equation}
    \gamma(\hat{\bm n})
    \equiv
    \gamma_1(\hat{\bm n})
    +
    i\gamma_2(\hat{\bm n}) .
\end{equation}
The spin-2 shear field is expanded {with spin-weighted spherical harmonics} as
\begin{equation}
    \gamma(\hat{\bm n})
    =
    \sum_{\ell m}
    \left(
    a^E_{\ell m}
    +
    i a^B_{\ell m}
    \right)
    {}_2Y_{\ell m}(\hat{\bm n}) .
    \label{eq:app_gamma_spin2_expansion}
\end{equation}
By the orthogonality of spin-weighted spherical harmonics, the inverse
relations are
\begin{align}
    a^E_{\ell m}
    &=
    \frac{1}{2}
    \int d\Omega\,
    \left[
    \gamma(\hat{\bm n})
    {}_2Y_{\ell m}^{*}(\hat{\bm n})
    +
    \gamma^*(\hat{\bm n})
    {}_{-2}Y_{\ell m}^{*}(\hat{\bm n})
    \right],
    \\
    a^B_{\ell m}
    &=
    \frac{1}{2i}
    \int d\Omega\,
    \left[
    \gamma(\hat{\bm n})
    {}_2Y_{\ell m}^{*}(\hat{\bm n})
    -
    \gamma^*(\hat{\bm n})
    {}_{-2}Y_{\ell m}^{*}(\hat{\bm n})
    \right].
\end{align}
In the presence of a survey window \(W(\hat{\bm n})\), we define the
windowed
{spherical harmonics coefficients}
as
\begin{align}
    \tilde a^E_{\ell m}
    &\equiv
    \frac{1}{2}
    \int d\Omega\,
    W(\hat{\bm n})
    \left[
    \gamma(\hat{\bm n})
    {}_2Y_{\ell m}^{*}(\hat{\bm n})
    +
    \gamma^*(\hat{\bm n})
    {}_{-2}Y_{\ell m}^{*}(\hat{\bm n})
    \right],
    \label{eq:app_atilde_E_def}
    \\
    \tilde a^B_{\ell m}
    &\equiv
    \frac{1}{2i}
    \int d\Omega\,
    W(\hat{\bm n})
    \left[
    \gamma(\hat{\bm n})
    {}_2Y_{\ell m}^{*}(\hat{\bm n})
    -
    \gamma^*(\hat{\bm n})
    {}_{-2}Y_{\ell m}^{*}(\hat{\bm n})
    \right].
    \label{eq:app_atilde_B_def}
\end{align}
The window-convolved power spectrum is then defined by
\begin{equation}
    \tilde C^{XY}_{\ell}
    \equiv
    \frac{1}{2\ell+1}
    \sum_{m=-\ell}^{\ell}
    \tilde a^X_{\ell m}
    \tilde a^{Y*}_{\ell m},
    \qquad
    X,Y\in\{E,B\}.
    \label{eq:app_pseudo_cl_def}
\end{equation}
Using Eqs.~\eqref{eq:app_atilde_E_def} and
\eqref{eq:app_atilde_B_def}, the four window-convolved power spectrum can be written explicitly as
\begin{align}
    \tilde C^{EE}_{\ell}
    &=
    \frac{1}{2\ell+1}
    \sum_{m=-\ell}^{\ell}
    \frac{1}{4}
    \int d\Omega_a d\Omega_b\,
    W(\hat{\bm n}_a)W(\hat{\bm n}_b)
    \nonumber\\
    &\quad \times
    \Bigl[
    \gamma(\hat{\bm n}_a)\gamma^*(\hat{\bm n}_b)
    {}_2Y_{\ell m}^{*}(\hat{\bm n}_a)
    {}_2Y_{\ell m}(\hat{\bm n}_b)
    \nonumber\\
    &\qquad
    +
    \gamma(\hat{\bm n}_a)\gamma(\hat{\bm n}_b)
    {}_2Y_{\ell m}^{*}(\hat{\bm n}_a)
    {}_{-2}Y_{\ell m}(\hat{\bm n}_b)
    \nonumber\\
    &\qquad
    +
    \gamma^*(\hat{\bm n}_a)\gamma^*(\hat{\bm n}_b)
    {}_{-2}Y_{\ell m}^{*}(\hat{\bm n}_a)
    {}_2Y_{\ell m}(\hat{\bm n}_b)
    \nonumber\\
    &\qquad
    +
    \gamma^*(\hat{\bm n}_a)\gamma(\hat{\bm n}_b)
    {}_{-2}Y_{\ell m}^{*}(\hat{\bm n}_a)
    {}_{-2}Y_{\ell m}(\hat{\bm n}_b)
    \Bigr],
    \label{eq:app_cl_EE_expanded}
    \end{align}

    \begin{align}
    \tilde C^{EB}_{\ell}
    &=
    \frac{1}{2\ell+1}
    \sum_{m=-\ell}^{\ell}
    \frac{-1}{4i}
    \int d\Omega_a d\Omega_b\,
    W(\hat{\bm n}_a)W(\hat{\bm n}_b)
    \nonumber\\
    &\quad \times
    \Bigl[
    \gamma(\hat{\bm n}_a)\gamma^*(\hat{\bm n}_b)
    {}_2Y_{\ell m}^{*}(\hat{\bm n}_a)
    {}_2Y_{\ell m}(\hat{\bm n}_b)
    \nonumber\\
    &\qquad
    -
    \gamma(\hat{\bm n}_a)\gamma(\hat{\bm n}_b)
    {}_2Y_{\ell m}^{*}(\hat{\bm n}_a)
    {}_{-2}Y_{\ell m}(\hat{\bm n}_b)
    \nonumber\\
    &\qquad
    +
    \gamma^*(\hat{\bm n}_a)\gamma^*(\hat{\bm n}_b)
    {}_{-2}Y_{\ell m}^{*}(\hat{\bm n}_a)
    {}_2Y_{\ell m}(\hat{\bm n}_b)
    \nonumber\\
    &\qquad
    -
    \gamma^*(\hat{\bm n}_a)\gamma(\hat{\bm n}_b)
    {}_{-2}Y_{\ell m}^{*}(\hat{\bm n}_a)
    {}_{-2}Y_{\ell m}(\hat{\bm n}_b)
    \Bigr],
    \label{eq:app_cl_EB_expanded}
    \end{align}

    \begin{align}
    \tilde C^{BE}_{\ell}
    &=
    \frac{1}{2\ell+1}
    \sum_{m=-\ell}^{\ell}
    \frac{1}{4i}
    \int d\Omega_a d\Omega_b\,
    W(\hat{\bm n}_a)W(\hat{\bm n}_b)
    \nonumber\\
    &\quad \times
    \Bigl[
    \gamma(\hat{\bm n}_a)\gamma^*(\hat{\bm n}_b)
    {}_2Y_{\ell m}^{*}(\hat{\bm n}_a)
    {}_2Y_{\ell m}(\hat{\bm n}_b)
    \nonumber\\
    &\qquad
    +
    \gamma(\hat{\bm n}_a)\gamma(\hat{\bm n}_b)
    {}_2Y_{\ell m}^{*}(\hat{\bm n}_a)
    {}_{-2}Y_{\ell m}(\hat{\bm n}_b)
    \nonumber\\
    &\qquad
    -
    \gamma^*(\hat{\bm n}_a)\gamma^*(\hat{\bm n}_b)
    {}_{-2}Y_{\ell m}^{*}(\hat{\bm n}_a)
    {}_2Y_{\ell m}(\hat{\bm n}_b)
    \nonumber\\
    &\qquad
    -
    \gamma^*(\hat{\bm n}_a)\gamma(\hat{\bm n}_b)
    {}_{-2}Y_{\ell m}^{*}(\hat{\bm n}_a)
    {}_{-2}Y_{\ell m}(\hat{\bm n}_b)
    \Bigr],
    \label{eq:app_cl_BE_expanded}
    \end{align}

    \begin{align}
    \tilde C^{BB}_{\ell}
    &=
    \frac{1}{2\ell+1}
    \sum_{m=-\ell}^{\ell}
    \frac{1}{4}
    \int d\Omega_a d\Omega_b\,
    W(\hat{\bm n}_a)W(\hat{\bm n}_b)
    \nonumber\\
    &\quad \times
    \Bigl[
    \gamma(\hat{\bm n}_a)\gamma^*(\hat{\bm n}_b)
    {}_2Y_{\ell m}^{*}(\hat{\bm n}_a)
    {}_2Y_{\ell m}(\hat{\bm n}_b)
    \nonumber\\
    &\qquad
    -
    \gamma(\hat{\bm n}_a)\gamma(\hat{\bm n}_b)
    {}_2Y_{\ell m}^{*}(\hat{\bm n}_a)
    {}_{-2}Y_{\ell m}(\hat{\bm n}_b)
    \nonumber\\
    &\qquad
    -
    \gamma^*(\hat{\bm n}_a)\gamma^*(\hat{\bm n}_b)
    {}_{-2}Y_{\ell m}^{*}(\hat{\bm n}_a)
    {}_2Y_{\ell m}(\hat{\bm n}_b)
    \nonumber\\
    &\qquad
    +
    \gamma^*(\hat{\bm n}_a)\gamma(\hat{\bm n}_b)
    {}_{-2}Y_{\ell m}^{*}(\hat{\bm n}_a)
    {}_{-2}Y_{\ell m}(\hat{\bm n}_b)
    \Bigr].
    \label{eq:app_cl_BB_expanded}
\end{align}

We now consider the binned estimator of the shear two-point correlation
function.  To avoid ambiguity, we distinguish the shear defined in a fixed
coordinate basis from the shear defined in the local basis of each pair.  We
write
\begin{equation}
    \gamma^{\rm f}(\hat{\bm n})
    \equiv
    \gamma(\hat{\bm n})
    =
    \gamma_1(\hat{\bm n})+i\gamma_2(\hat{\bm n}) .
\end{equation}
If the local basis associated with the pair is rotated by an angle
\(\varphi\) with respect to the fixed basis, the locally defined shear is
\begin{equation}
    \gamma^{\rm loc}(\hat{\bm n};\varphi)
    \equiv
    \gamma_t(\hat{\bm n})+i\gamma_\times(\hat{\bm n})
    =
    e^{-2i\varphi}
    \gamma^{\rm f}(\hat{\bm n}) .
    \label{eq:app_local_shear_def}
\end{equation}
Therefore,
\begin{align}
    \gamma^{\rm loc}(\hat{\bm n}_a;\varphi_a)
    \gamma^{{\rm loc}*}(\hat{\bm n}_b;\varphi_b)
    &=
    e^{-2i\varphi_a}
    e^{+2i\varphi_b}
    \gamma^{\rm f}(\hat{\bm n}_a)
    \gamma^{{\rm f}*}(\hat{\bm n}_b),
    \label{eq:app_local_product_plus}
    \\
    \gamma^{\rm loc}(\hat{\bm n}_a;\varphi_a)
    \gamma^{\rm loc}(\hat{\bm n}_b;\varphi_b)
    &=
    e^{-2i\varphi_a}
    e^{-2i\varphi_b}
    \gamma^{\rm f}(\hat{\bm n}_a)
    \gamma^{\rm f}(\hat{\bm n}_b).
    \label{eq:app_local_product_minus}
\end{align}
For \(\xi_+\), the binned estimator is written as
\begin{align}
    \hat\xi_+[\theta_-,\theta_+]
    \frac{N_{\rm pair}[\theta_-,\theta_+]}{n^2}
    &=
    \int_{\cos\theta_+}^{\cos\theta_-}
    d\cos\theta
    \int d\Omega_a d\Omega_b\,
    W(\hat{\bm n}_a)W(\hat{\bm n}_b)
    \nonumber\\
    &\quad \times
    {\rm Re}
    \left\{
    \delta_{\rm D}
    \left(
    \cos\theta_{ab}-\cos\theta
    \right)
    \gamma^{\rm loc}(\hat{\bm n}_a;\varphi_a)
    \gamma^{{\rm loc}*}(\hat{\bm n}_b;\varphi_b)
    \right\}.
    \label{eq:app_xip_start}
\end{align}
Using Eq.~\eqref{eq:app_local_product_plus}, this becomes
\begin{align}
    \hat\xi_+[\theta_-,\theta_+]
    \frac{N_{\rm pair}[\theta_-,\theta_+]}{n^2}
    &=
    \int_{\cos\theta_+}^{\cos\theta_-}
    d\cos\theta
    \int d\Omega_a d\Omega_b\,
    W(\hat{\bm n}_a)W(\hat{\bm n}_b)
    \nonumber\\
    &\quad \times
    {\rm Re}
    \left\{
    \delta_{\rm D}
    \left(
    \cos\theta_{ab}-\cos\theta
    \right)
    e^{-2i\varphi_a}
    e^{+2i\varphi_b}
    \gamma^{\rm f}(\hat{\bm n}_a)
    \gamma^{{\rm f}*}(\hat{\bm n}_b)
    \right\}.
    \label{eq:app_xip_fixed_basis}
\end{align}
Substituting Eq.~\eqref{eq:app_delta_spinY} with \(s=s'=2\), and choosing
the Euler-angle phases consistently with the local-basis rotations in
Eq.~\eqref{eq:app_local_product_plus}, we obtain
\begin{align}
    \hat\xi_+[\theta_-,\theta_+]
    \frac{N_{\rm pair}[\theta_-,\theta_+]}{n^2}
    &=
    2\pi
    \sum_{\ell m}
    \int_{\cos\theta_+}^{\cos\theta_-}
    d\cos\theta\,
    d^\ell_{22}(\theta)
    \nonumber\\
    &\quad \times
    {\rm Re}
    \left\{
    \int d\Omega_a d\Omega_b\,
    W(\hat{\bm n}_a)W(\hat{\bm n}_b)
    \gamma^{\rm f}(\hat{\bm n}_a)
    \gamma^{{\rm f}*}(\hat{\bm n}_b)
    {}_2Y_{\ell m}^{*}(\hat{\bm n}_a)
    {}_2Y_{\ell m}(\hat{\bm n}_b)
    \right\}.
    \label{eq:app_xip_spin22_term}
\end{align}
The term inside the real part is precisely the spin-\((2,2)\) term appearing
in Eqs.~\eqref{eq:app_cl_EE_expanded} and
\eqref{eq:app_cl_BB_expanded}.  The terms with spin weights
\((2,-2)\) and \((-2,2)\) cancel between
\(\tilde C_\ell^{EE}\) and \(\tilde C_\ell^{BB}\).  Hence
\begin{equation}
    \frac{1}{2\ell+1}
    \sum_m
    {\rm Re}
    \left[
    \int d\Omega_a d\Omega_b\,
    W_aW_b\,
    \gamma^{\rm f}_a
    \gamma^{{\rm f}*}_b
    {}_2Y_{\ell m}^{*}(\hat{\bm n}_a)
    {}_2Y_{\ell m}(\hat{\bm n}_b)
    \right]
    =
    \tilde C_\ell^{EE}
    +
    \tilde C_\ell^{BB},
    \label{eq:app_xip_cl_identity}
\end{equation}
where \(W_a\equiv W(\hat{\bm n}_a)\), \(W_b\equiv W(\hat{\bm n}_b)\),
\(\gamma^{\rm f}_a\equiv\gamma^{\rm f}(\hat{\bm n}_a)\), and
\(\gamma^{\rm f}_b\equiv\gamma^{\rm f}(\hat{\bm n}_b)\).  Therefore,
\begin{align}
    \hat\xi_+[\theta_-,\theta_+]
    \frac{N_{\rm pair}[\theta_-,\theta_+]}{n^2}
    &=
    2\pi
    \sum_{\ell}
    (2\ell+1)
    \left(
    \tilde C^{EE}_{\ell}
    +
    \tilde C^{BB}_{\ell}
    \right)
    \int_{\cos\theta_+}^{\cos\theta_-}
    d\cos\theta\,
    d^\ell_{22}(\theta).
    \label{eq:app_xip_pseudo_cl}
\end{align}
Similarly, for \(\xi_-\), we start from
\begin{align}
    \hat\xi_-[\theta_-,\theta_+]
    \frac{N_{\rm pair}[\theta_-,\theta_+]}{n^2}
    &=
    \int_{\cos\theta_+}^{\cos\theta_-}
    d\cos\theta
    \int d\Omega_a d\Omega_b\,
    W(\hat{\bm n}_a)W(\hat{\bm n}_b)
    \nonumber\\
    &\quad \times
    {\rm Re}
    \left\{
    \delta_{\rm D}
    \left(
    \cos\theta_{ab}-\cos\theta
    \right)
    \gamma(\hat{\bm n}_a)
    \gamma(\hat{\bm n}_b)
    \right\}.
    \label{eq:app_xim_start}
\end{align}
Using Eq.~\eqref{eq:app_delta_spinY} with \(s=2\) and \(s'=-2\), we obtain
\begin{align}
    \hat\xi_-[\theta_-,\theta_+]
    \frac{N_{\rm pair}[\theta_-,\theta_+]}{n^2}
    &=
    \int_{\cos\theta_+}^{\cos\theta_-}
    d\cos\theta
    \int d\Omega_a d\Omega_b\,
    W(\hat{\bm n}_a)W(\hat{\bm n}_b)
    \nonumber\\
    &\quad \times
    {\rm Re}
    \left\{
    2\pi
    \sum_{\ell m}
    d^\ell_{2,-2}(\theta)
    e^{2i\alpha}
    e^{-2i\gamma}
    {}_2Y_{\ell m}^{*}(\hat{\bm n}_a)
    {}_{-2}Y_{\ell m}(\hat{\bm n}_b)
    \gamma(\hat{\bm n}_a)
    \gamma(\hat{\bm n}_b)
    \right\}.
    \label{eq:app_xim_after_delta}
\end{align}
The spin combination in Eq.~\eqref{eq:app_xim_after_delta} corresponds to the
terms with opposite spin weights in the window-convolved power spectrum.  Taking the real part
and using Eqs.~\eqref{eq:app_cl_EE_expanded}--\eqref{eq:app_cl_BB_expanded},
we find
\begin{align}
    \hat\xi_-[\theta_-,\theta_+]
    \frac{N_{\rm pair}[\theta_-,\theta_+]}{n^2}
    &=
    2\pi
    \sum_{\ell}
    (2\ell+1)
    \left(
    \tilde C^{EE}_{\ell}
    -
    \tilde C^{BB}_{\ell}
    \right)
    \int_{\cos\theta_+}^{\cos\theta_-}
    d\cos\theta\,
    d^\ell_{2,-2}(\theta).
    \label{eq:app_xim_pseudo_cl}
\end{align}
The terms proportional to \(\tilde C_\ell^{EB}\) and \(\tilde C_\ell^{BE}\)
are purely imaginary in the above real-part operation and do not contribute to
\(\xi_\pm\) for the parity-symmetric case considered here.
Finally, dividing by \(N_{\rm pair}[\theta_-,\theta_+]/n^2\), the binned
relations are
\begin{align}
    \hat\xi_+[\theta_-,\theta_+]
    &=
    \frac{2\pi n^2}{N_{\rm pair}[\theta_-,\theta_+]}
    \sum_{\ell}
    (2\ell+1)
    \left(
    \tilde C^{EE}_{\ell}
    +
    \tilde C^{BB}_{\ell}
    \right)
    \int_{\cos\theta_+}^{\cos\theta_-}
    d\cos\theta\,
    d^\ell_{22}(\theta),
    \label{eq:app_xip_final}
    \\
    \hat\xi_-[\theta_-,\theta_+]
    &=
    \frac{2\pi n^2}{N_{\rm pair}[\theta_-,\theta_+]}
    \sum_{\ell}
    (2\ell+1)
    \left(
    \tilde C^{EE}_{\ell}
    -
    \tilde C^{BB}_{\ell}
    \right)
    \int_{\cos\theta_+}^{\cos\theta_-}
    d\cos\theta\,
    d^\ell_{2,-2}(\theta).
    \label{eq:app_xim_final}
\end{align}

\subsection{Covariance expression}
\label{app:covariance_expression}

We then derive the covariance matrix of the binned shear
two-point correlation functions from the covariance matrix of the
window-convolved pseudo-$C_\ell$ power spectra.  Since the estimators
derived in Appendix~\ref{app:xi_pseudo_cl_relation} are linear transformations of the pseudo-$C_\ell$,
their covariance matrices can also be obtained by a linear projection.

From Eqs.~(A.29) and (A.30), we write the binned correlation functions as
\begin{align}
\hat{\xi}_{+,i}
&=
\sum_{\ell}
P^{+}_{i\ell}
\left(
\tilde{C}^{\rm EE}_{\ell}
+
\tilde{C}^{\rm BB}_{\ell}
\right),
\\
\hat{\xi}_{-,i}
&=
\sum_{\ell}
P^{-}_{i\ell}
\left(
\tilde{C}^{\rm EE}_{\ell}
-
\tilde{C}^{\rm BB}_{\ell}
\right),
\end{align}
where $i$ denotes the angular bin and the projection matrices are defined as
\begin{align}
P^{+}_{i\ell}
&\equiv
\frac{2\pi n^2}{N_{{\rm pair},i}}
(2\ell+1)
\int_{\cos\theta_{i,+}}^{\cos\theta_{i,-}}
d\cos\theta\,
d^{\ell}_{22}(\theta),
\\
P^{-}_{i\ell}
&\equiv
\frac{2\pi n^2}{N_{{\rm pair},i}}
(2\ell+1)
\int_{\cos\theta_{i,+}}^{\cos\theta_{i,-}}
d\cos\theta\,
d^{\ell}_{2,-2}(\theta).
\end{align}
Here, $\theta_{i,-}$ and $\theta_{i,+}$ are the lower and upper edges of
the $i$-th angular bin, respectively.

We define the covariance matrix of the window-convolved power spectra as
\begin{equation}
{\rm Cov}^{XY,ZW}_{\ell\ell'}
\equiv
\left\langle
\tilde{C}^{XY}_{\ell}
\tilde{C}^{ZW}_{\ell'}
\right\rangle 
-
\left\langle
\tilde{C}^{XY}_{\ell}
\right\rangle
\left\langle
\tilde{C}^{ZW}_{\ell'}
\right\rangle.
\end{equation}
The covariance matrix of $\xi_+$ is then given by
\begin{align}
\label{eq:app_xip_cov}
{\rm Cov}^{++}_{ij}
&=
\sum_{\ell\ell'}
P^{+}_{i\ell}
P^{+}_{j\ell'}
\left[
{\rm Cov}^{\rm EE,EE}_{\ell\ell'}
+
{\rm Cov}^{\rm EE,BB}_{\ell\ell'}
+
{\rm Cov}^{\rm BB,EE}_{\ell\ell'}
+
{\rm Cov}^{\rm BB,BB}_{\ell\ell'}
\right].
\end{align}
Similarly, the covariance matrix of $\xi_-$ is
\begin{align}
\label{eq:app_xim_cov}
{\rm Cov}^{--}_{ij}
&=
\sum_{\ell\ell'}
P^{-}_{i\ell}
P^{-}_{j\ell'}
\left[
{\rm Cov}^{\rm EE,EE}_{\ell\ell'}
-
{\rm Cov}^{\rm EE,BB}_{\ell\ell'}
-
{\rm Cov}^{\rm BB,EE}_{\ell\ell'}
+
{\rm Cov}^{\rm BB,BB}_{\ell\ell'}
\right].
\end{align}

Thus, the covariance matrix of the shear two-point correlation functions
can be obtained by projecting the covariance matrices of the window-convolved $C_\ell$.

In Appendix~B, we validate this prediction by comparing
it with Gaussian simulations of spin-2 shear fields.

\section{Results for shear}
\label{app:shear_result}

In this appendix, we validate the formalism
presented in Appendix~A.
{First we construct Gaussian simulations of a spin-2 shear field following Sec.~\ref{sec:simulations}. When generating shear field, we specify the input power spectra, $\{C_{\ell}^{EE}, C_{\ell}^{EB}=0,C_{\ell}^{BB}=0 \}$, where the input $E$-mode power spectrum $C_{\ell}^{EE}$ is the same as the input convergence power spectrum in Sec.~\ref{sec:simulations}.
}
We compare the two-point correlation functions and their
covariances measured from masked Gaussian maps with the analytical
predictions constructed from the window-convolved power
spectra {(Eqs.~\ref{eq:app_xip_cov} and \ref{eq:app_xim_cov})}.

\subsection{Signal of $\tilde{C}_\ell$ and $\xi_\pm(\theta)$}
\label{app:shear_xi}

We first test whether 
{the projection of the window-convolved power spectra}
described in Appendix~\ref{app:xi_pseudo_cl_relation}
correctly produces the spin-2 correlation functions $\xi_\pm(\theta)$.

\begin{figure}
\centering

\begin{minipage}[t]{.48\textwidth}
\vspace{0pt}
\centering
\includegraphics[width=\linewidth]{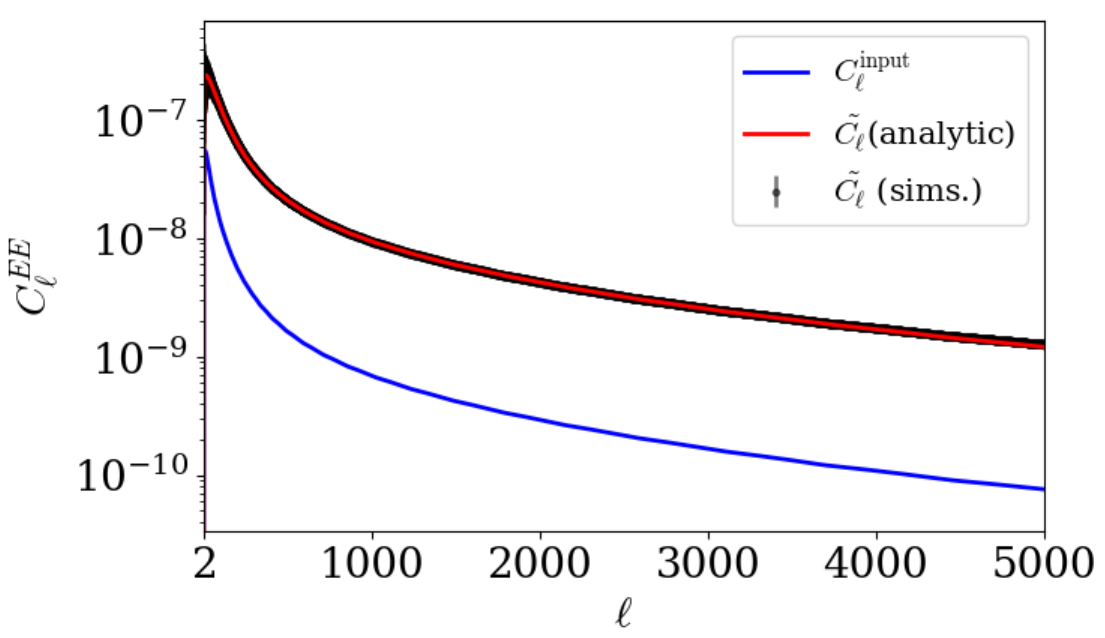}
\caption{Comparison of the $E$-mode power spectrum measured from Gaussian simulations and the analytical prediction derived from the pseudo-$C_\ell$ formalism.
The blue solid line shows the input power spectrum.}
\label{fig:clee}
\end{minipage}
\hfill
\begin{minipage}[t]{.48\textwidth}
\vspace{0pt}
\centering
\includegraphics[width=\linewidth]{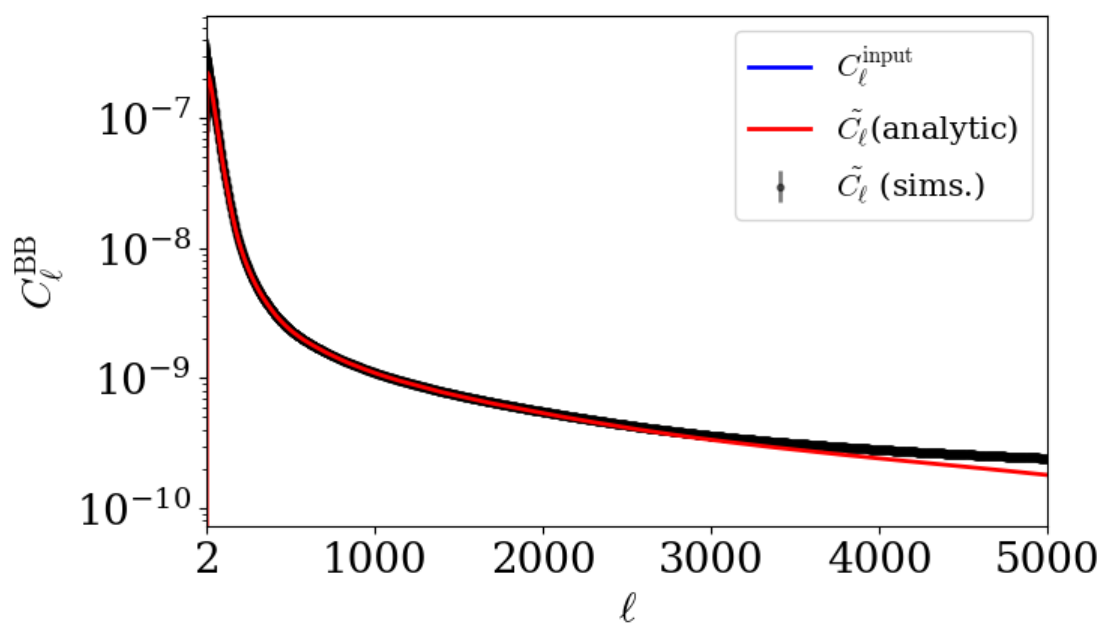}
\caption{Comparison of the $B$-mode power spectrum measured from Gaussian simulations and the analytical prediction derived from the pseudo-$C_\ell$ formalism.
{Note that the input $B$-mode power spectrum is zero and measured $B$-mode is leakage from $E$-mode due to mode-coupling.}
}
\label{fig:clbb}
\end{minipage}

\end{figure}

Figures~\ref{fig:clee} and~\ref{fig:clbb} compare the analytical predictions of the $E$-mode and
$B$-mode power spectra with those measured from the simulations.
The analytically calculated $E$-mode power spectrum agrees well with the
direct measurement over the full angular range.

\begin{figure}[H]
\centering

\begin{minipage}[t]{.48\textwidth}
\vspace{0pt}
\centering
\includegraphics[width=\linewidth]{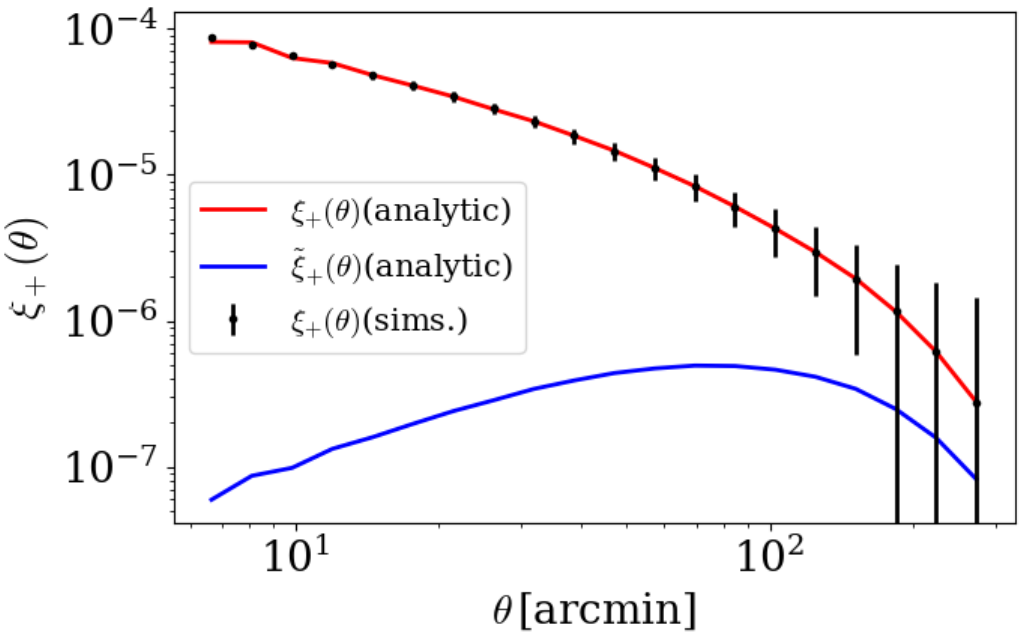}
\caption{Comparison of the shear two-point correlation function
$\xi_{+}(\theta)$ reconstructed from the analytical
window-convolved power spectra and that measured directly from
Gaussian simulations.
Excellent agreement is obtained over the full angular range.}
\label{fig:xip}
\end{minipage}
\hfill
\begin{minipage}[t]{.48\textwidth}
\vspace{0pt}
\centering
\includegraphics[width=\linewidth]{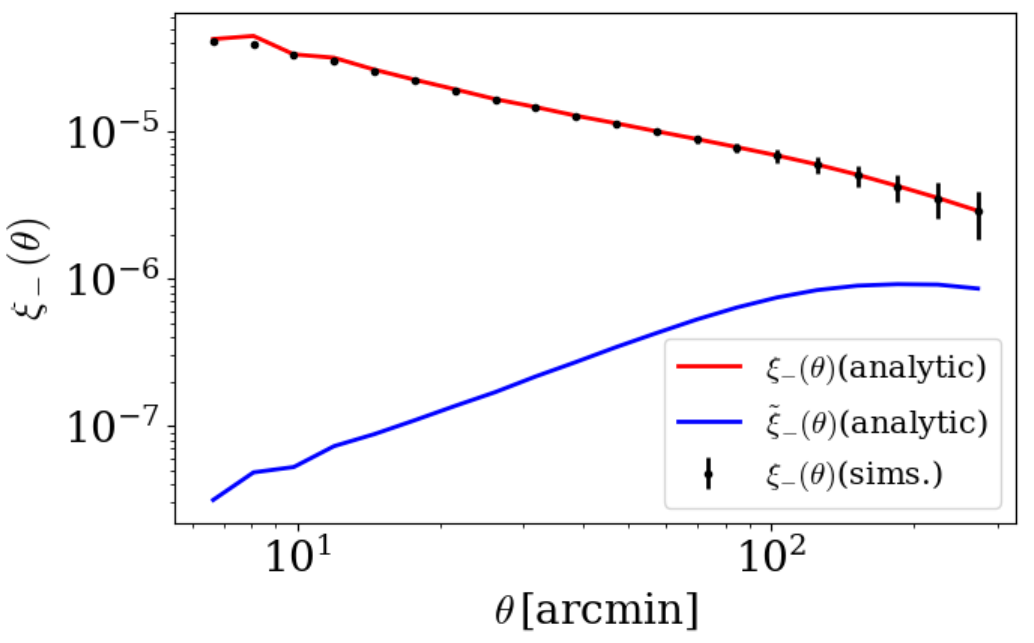}
\caption{Comparison of the shear two-point correlation functions reconstructed from the analytical window-convolved power spectra and those measured directly from Gaussian simulations.}
\label{fig:xim}
\end{minipage}

\end{figure}

Figures~\ref{fig:xip} and~\ref{fig:xim} compare the corresponding shear two-point correlation
functions, $\xi_{+}$ and $\xi_{-}$, 
{calculated as a projection of window-convolved $C_{\ell}$ (Eqs.~\ref{eq:app_xip_pseudo_cl} and \ref{eq:app_xim_pseudo_cl})}
with those measured directly from the simulations.
We find a good agreement {and therefore confirm Eqs.~\ref{eq:app_xip_pseudo_cl} and \ref{eq:app_xim_pseudo_cl} work well.}

\subsection{Covariance}
\label{app:shear_cov}

We next test whether the $\xi_{\pm}$ covariance predicted by
the pseudo-$C_\ell$ formalism reproduces the covariance
measured from simulations.
We reconstruct the covariance matrices of
$\xi_{\pm}(\theta)$  through the transformation derived in Appendix~A.2.

Figures~\ref{fig:cov_ee_ee} and~\ref{fig:cov_bb_bb} compare the diagonal elements of the analytical Gaussian covariance for the $E$-mode and $B$-mode power spectra with those measured from the simulations.
The analytical predictions reproduce the simulated diagonal auto-covariances well over the multipole range considered.

\begin{figure}[H]
\centering
\begin{minipage}[H]{.60\textwidth}
\centering
\includegraphics[width=\linewidth]{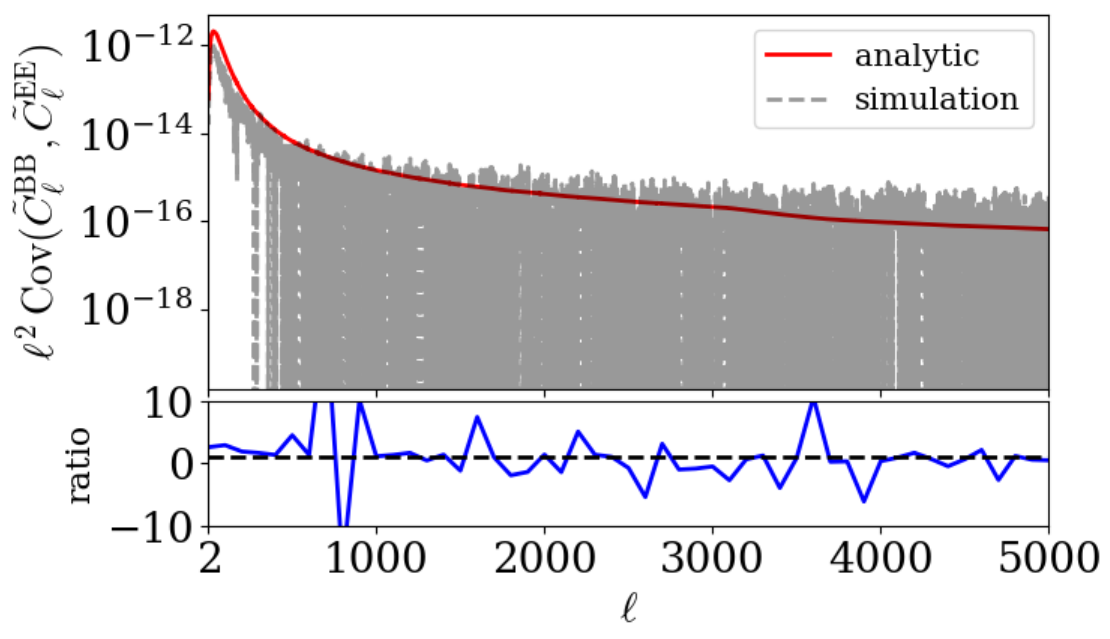}
\caption{Comparison of the analytic Gaussian cross-covariance between the $E$-mode and $B$-mode power spectra
with that measured from Gaussian simulations. 
The horizontal line in the lower panel denotes ${\rm ratio}=1$.
The simulated cross-covariance exhibits an oscillatory structure that is not reproduced by the analytic calculation.}
\label{fig:cov_ee_bb}
\end{minipage}
\end{figure}

Figure~\ref{fig:cov_ee_bb} shows the diagonal elements of the
cross-covariance between the $E$-mode and $B$-mode power spectra.
In contrast to the auto-covariances, the simulation result exhibits an
oscillatory structure that is not reproduced by the analytical
calculation.

\begin{figure}[H]
    \centering
    \includegraphics[width=0.8\linewidth]{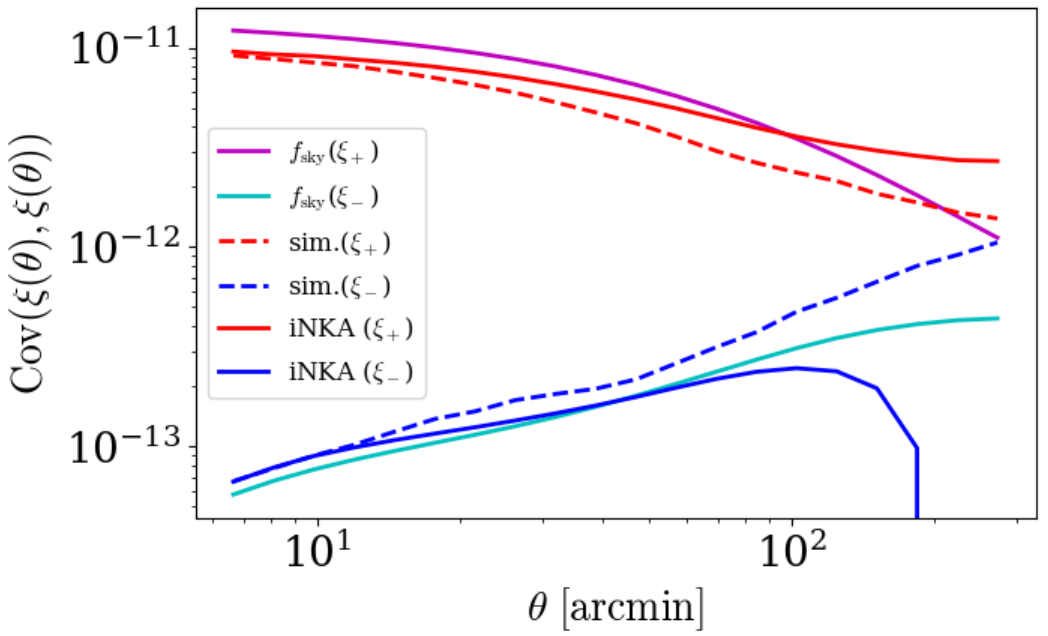}
    \caption{Comparison of the analytic Gaussian covariance of the two-point correlation functions with simulations.
    The diagonal covariance, $\mathrm{Cov}(\xi_\pm,\xi_\pm)$,
    is shown as a function of angular separation.
    The red and blue solid lines show the analytic prediction reconstructed from the harmonic-space covariance using the transformation derived in Appendix~A (iNKA), the dashed lines show the covariance measured from Gaussian simulations, and the magenta and cyan solid lines show the prediction based on the $f_{\rm sky}$ approximation.
    }
    \label{fig:cov_xipm}
\end{figure}

\begin{figure}[H]
\centering

\begin{minipage}[t]{.48\textwidth}
\vspace{0pt}
\centering
\includegraphics[width=\linewidth]{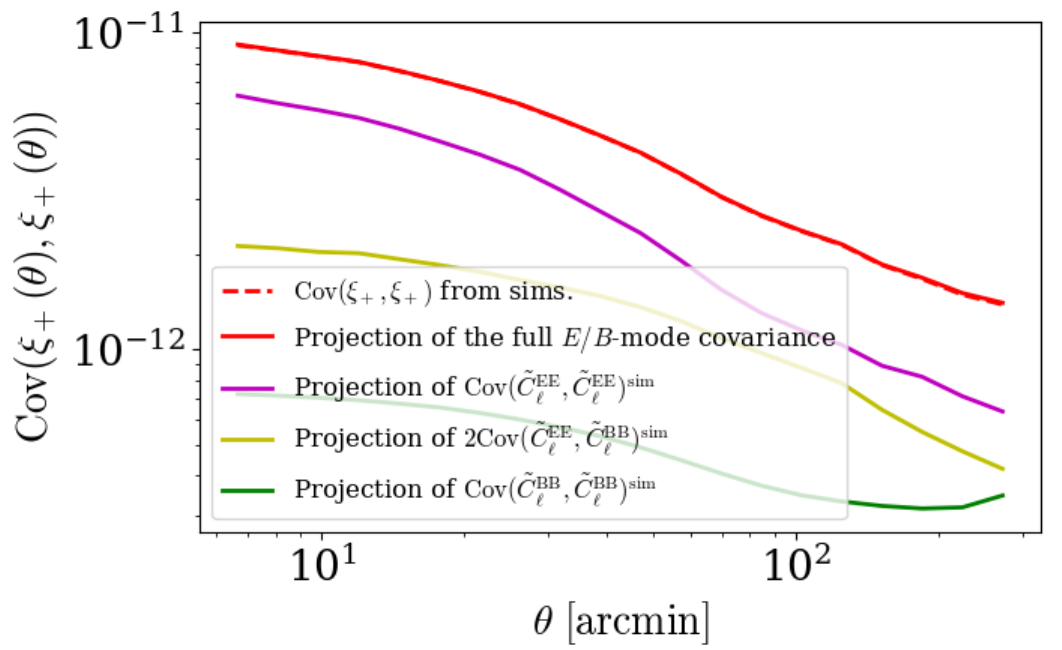}
\caption{Decomposition of the diagonal covariance of
$\xi_{+}(\theta)$.
The dashed red line shows the covariance measured from Gaussian
simulations, while the solid red line shows the covariance obtained by
projecting the full harmonic-space covariance of Gaussian simulations.
The magenta, green, and yellow lines denote the projected contributions from the E-mode auto-covariances, B-mode auto-covariance, and EB cross-covariance, respectively.
}
\label{fig:covxip}
\end{minipage}
\hfill
\begin{minipage}[t]{.48\textwidth}
\vspace{0pt}
\centering
\includegraphics[width=\linewidth]{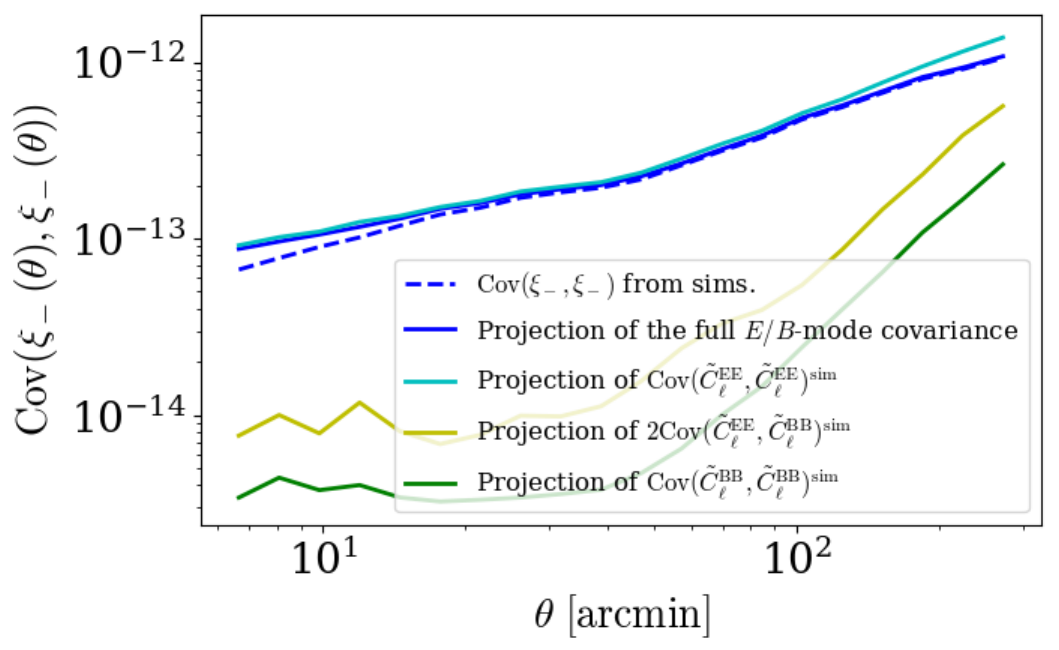}
\caption{The dashed blue line shows the covariance measured from Gaussian
simulations, while the solid blue line shows the covariance obtained by
projecting the full harmonic-space covariance of Gaussian simulations.
The cyan, green, and yellow lines denote the projected contributions from the E-mode auto-covariances, B-mode auto-covariance, and EB cross-covariance, respectively. 
For visualization purposes, the solid yellow line shows the projected term corresponding to $+2{\rm Cov}(\tilde{C}_\ell^{\rm EE},\tilde{C}_\ell^{\rm BB})^{\rm sim}$.
In the full covariance of $\xi_{-}(\theta)$, however, this term enters with the opposite sign, as $-2{\rm Cov}(\tilde{C}_\ell^{\rm EE},\tilde{C}_\ell^{\rm BB})^{\rm sim}$.}

\label{fig:covxim}
\end{minipage}

\end{figure}

Figure~\ref{fig:cov_xipm} compares the diagonal covariance of the shear two-point correlation functions reconstructed from the harmonic-space covariance with that measured directly from the simulations.
The figure also includes the prediction based on the $f_{\rm sky}$ approximation.
Although the diagonal $EE$ and $BB$ auto-covariances are individually well reproduced, the projected analytical covariance of $\xi_{\pm}(\theta)$ does not reproduce the simulation result with the same accuracy.
{Figure~\ref{fig:covxip}, and ~\ref{fig:covxim} indicates that the oscillatory $EE$--$BB$ cross-covariance contains a genuine contribution required for the projection to configuration space.}

\begin{figure}[H]
    \centering
    \includegraphics[width=0.8\linewidth]{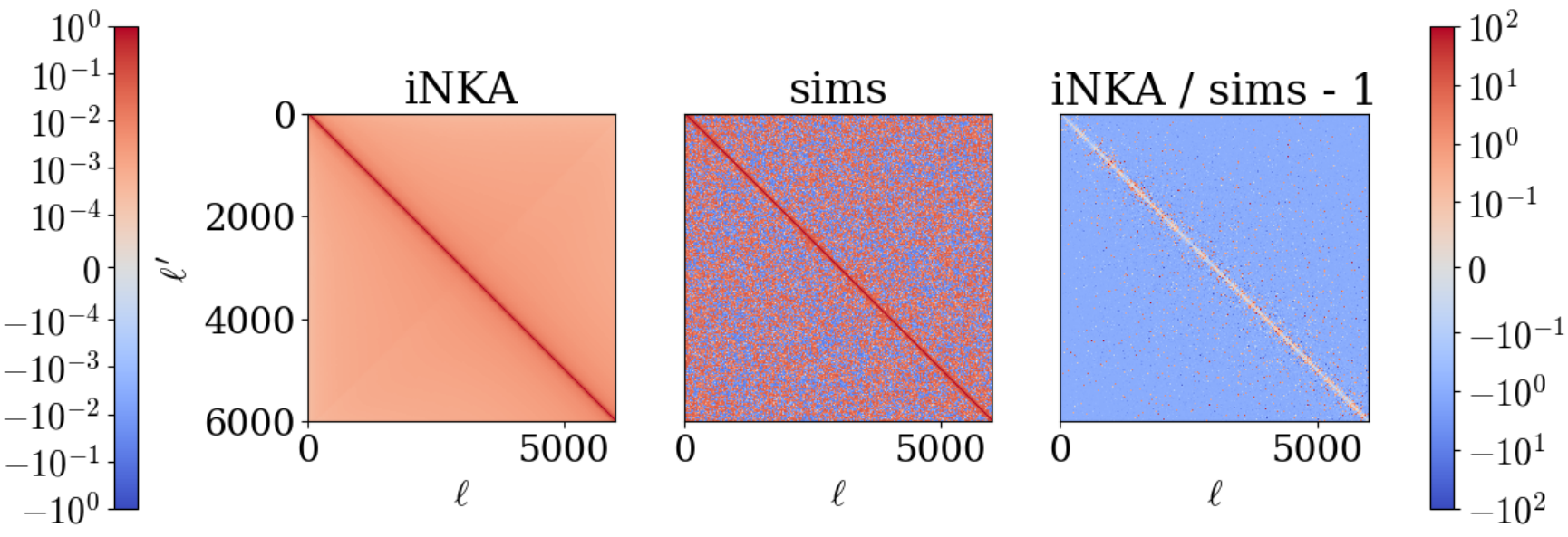}
    \caption{Correlation matrices of the $E$-mode covariance,
    $\mathrm{Cov}(\tilde{C}^{EE}_{\ell},\tilde{C}^{EE}_{\ell'})$,
computed using the iNKA (left) and measured from Gaussian simulations (center).
{The colors corresponding to the map values are depicted on the left.
The right plot shows the fractional difference $\mathrm{Cov}(\tilde{C}^{EE}_\ell,\tilde{C}^{EE}_{\ell'})^{\rm iNKA} / \mathrm{Cov}(\tilde{C}^{EE}_\ell,\tilde{C}^{EE}_{\ell'})^{\rm sim} - 1$ and color bar is shown on the right.}
The iNKA reproduces the diagonal structure accurately,
while noticeable differences remain in the off-diagonal components.}
\label{fig:matrix_EE_EE}
\end{figure}

\begin{figure}[H]
    \centering
    \includegraphics[width=0.8\linewidth]{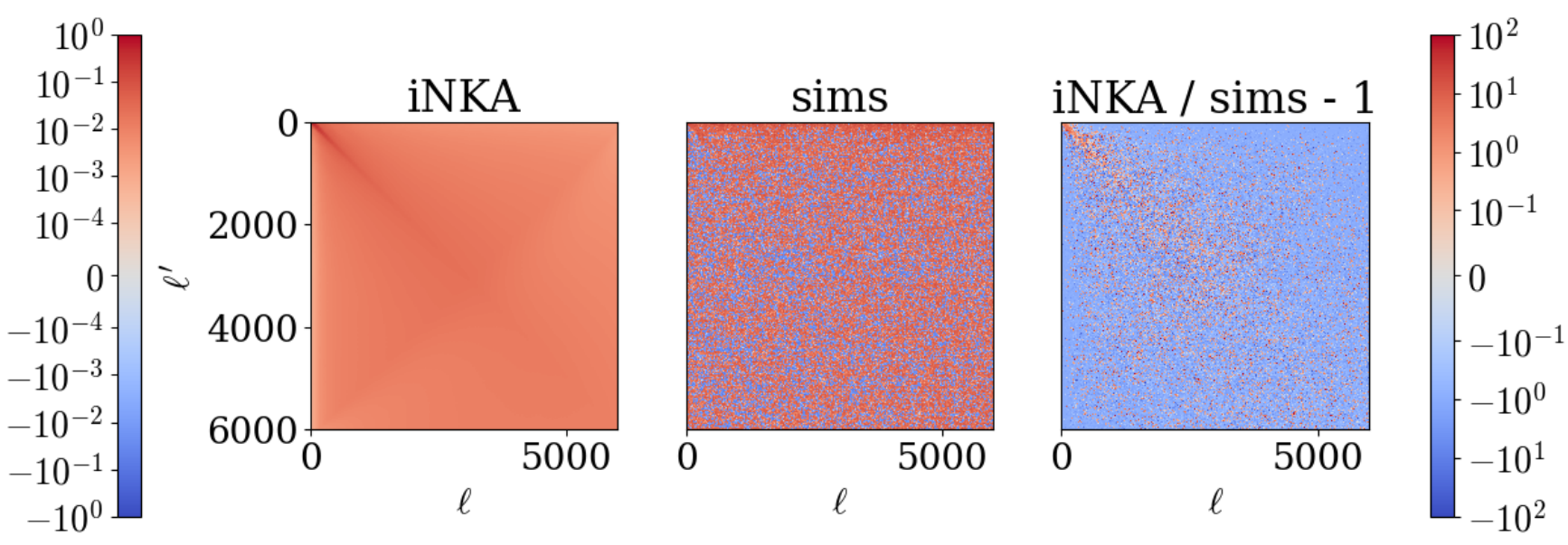}
    \caption{Similar to Fig.~\ref{fig:matrix_EE_EE}, but for the cross-covariance,
    $\mathrm{Cov}(\tilde{C}^{EE}_{\ell},\tilde{C}^{BB}_{\ell'})$.}
    \label{fig:matrix_EE_BB}
\end{figure}

\begin{figure}[H]
    \centering
    \includegraphics[width=0.8\linewidth]{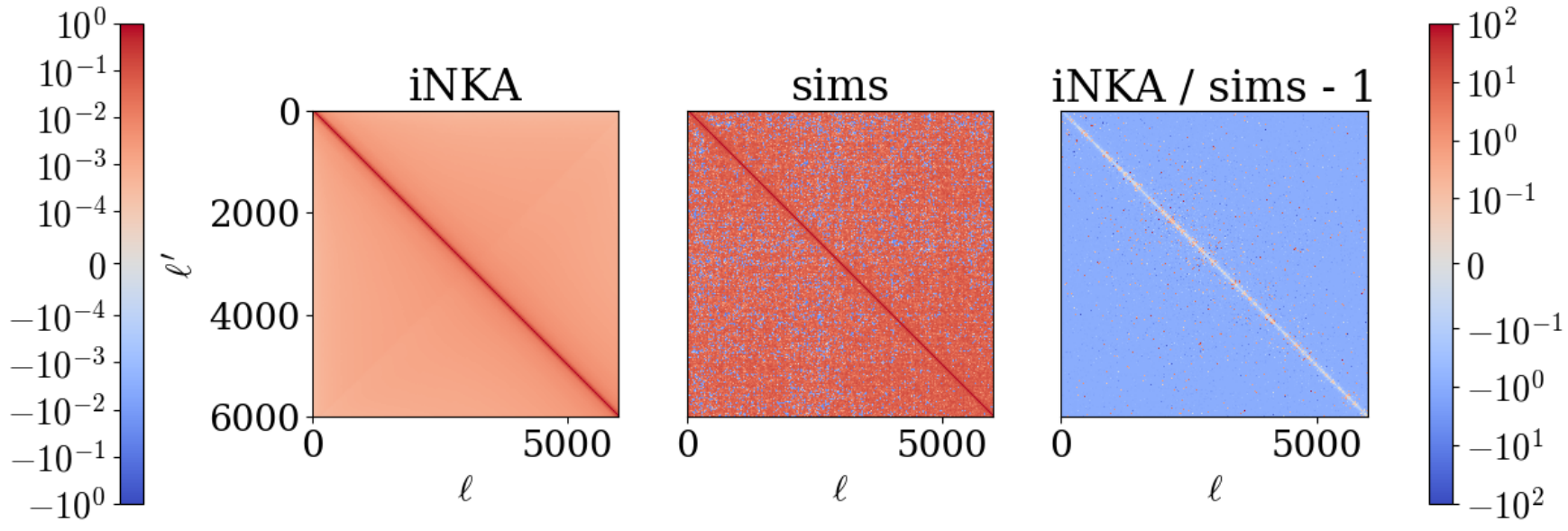}
    \caption{Similar to Fig.~\ref{fig:matrix_EE_EE}, but for the $B$-mode covariance,
    $\mathrm{Cov}(\tilde{C}^{BB}_{\ell},\tilde{C}^{BB}_{\ell'})$.
    }
    \label{fig:matrix_BB_BB}
\end{figure}

Figures~\ref{fig:matrix_EE_EE}, ~\ref{fig:matrix_EE_BB}, and ~\ref{fig:matrix_BB_BB} compare the correlation coefficient matrices of the $EE$ auto-covariance, the $EE$-$BB$ cross-covariance, and the $BB$ auto-covariance, respectively.
As discussed in the main text for a convergence field, this result
suggests that discrepancies in the off-diagonal elements of the
harmonic-space covariance contribute to the reduced accuracy of
the projection to configuration space.


\begin{figure}[H]
\centering
\begin{minipage}[H]{.48\textwidth}
\vspace{0pt}
\centering
\includegraphics[width=\linewidth]{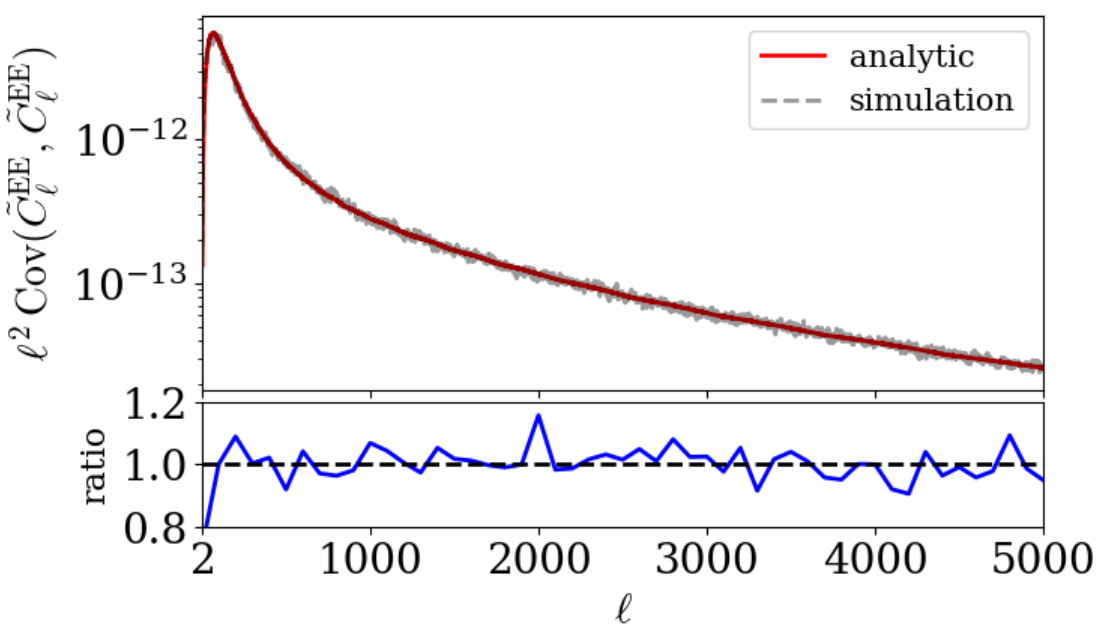}
\caption{Comparison of the analytic Gaussian covariance of the $E$-mode power spectrum with simulations.
    The upper panel shows the diagonal elements of the covariance matrix, while the lower panel presents the ratio of the analytic prediction to the simulation.
    The analytic calculation accurately reproduces the covariance measured from the simulations over the full multipole range.}
\label{fig:cov_ee_ee}
\end{minipage}
\hfill
\begin{minipage}[H]{.48\textwidth}
\vspace{0pt}
\centering
\includegraphics[width=\linewidth]{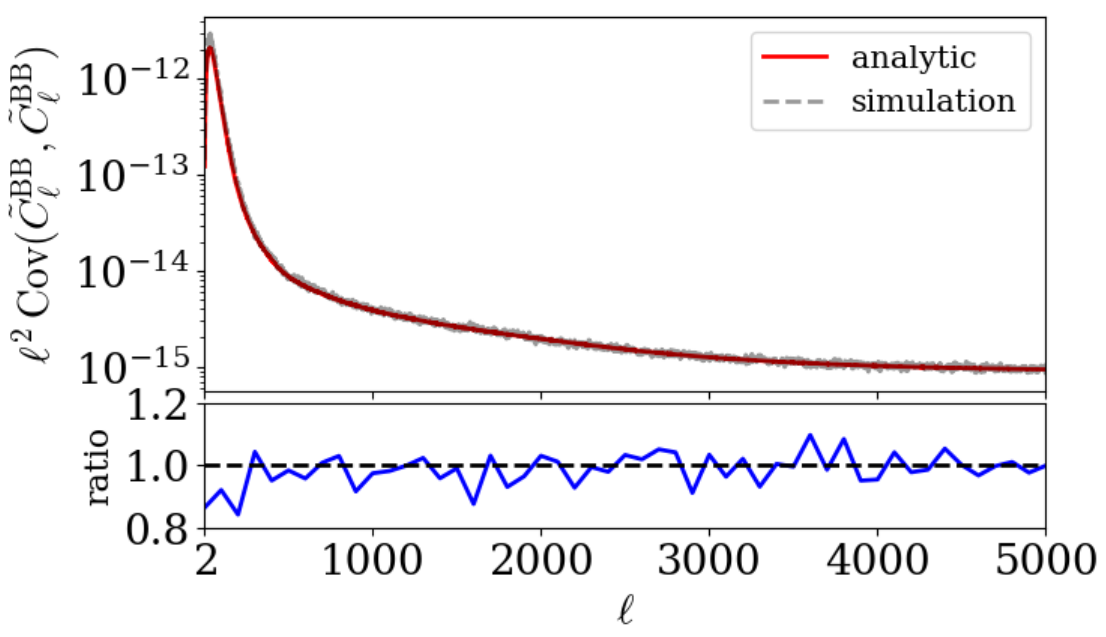}
\caption{Comparison of the analytic Gaussian covariance of the $B$-mode power spectrum with simulations.
The upper panel shows the diagonal elements of the covariance matrix,
while the lower panel presents the ratio of the analytic prediction to the simulation.}
\label{fig:cov_bb_bb}
\end{minipage}
\end{figure}


\acknowledgments
We thank Sunao Sugiyama and LSST DESC MCPCov group for useful discussion.
This work was supported in part by
JSPS KAKENHI Grant Number 20H05600, 20H05855, 23KJ0747, 24H00215, and by World Premier International Research Center Initiative (WPI Initiative), MEXT, Japan.

\bibliographystyle{JHEP}
\bibliography{refs}
\end{document}